\documentclass[pra,aps,reprint,a4paper,superscriptaddress,floatfix]{revtex4-2}
\usepackage{times}
\usepackage{graphicx}
\usepackage{epsfig}
\usepackage{amsfonts}
\usepackage{amsmath}
\usepackage{amssymb}
\usepackage{dsfont}
\usepackage{amsthm}
\usepackage{color,xcolor,ulem}
\usepackage[colorlinks=true,linkcolor=blue,citecolor=blue,urlcolor=blue,breaklinks]{hyperref}
\usepackage{braket}
\usepackage{physics}
\usepackage[breaklinks]{hyperref}

\definecolor{pinocolor}{HTML}{119911}

\begin{document}
	
	
\title{Energy-time and time-bin entanglement: past, present and future}
	
	
\author{Guilherme B. Xavier}
	\affiliation{Institutionen f\"{o}r Systemteknik, Link\"opings Universitet, 581 83 Link\"oping, Sweden}
 
 \author{Jan-\AA ke Larsson}
\affiliation{Institutionen f\"{o}r Systemteknik, Link\"opings Universitet, 581 83 Link\"oping, Sweden}

\author{Paolo Villoresi}
 \affiliation{Istituto Nazionale di Fisica Nucleare (INFN)—Sezione di Padova, Via Marzolo 8, 35131, Padova, Italy }
 \affiliation{Dipartimento di Ingegneria dell'Informazione, Universit\`a degli Studi di Padova, via Gradenigo 6B, 35131 Padova, Italy}

 \author{Giuseppe Vallone}
 \affiliation{Dipartimento di Ingegneria dell'Informazione, Universit\`a degli Studi di Padova, via Gradenigo 6B, 35131 Padova, Italy}
 \affiliation{Istituto Nazionale di Fisica Nucleare (INFN)—Sezione di Padova, Via Marzolo 8, 35131, Padova, Italy }
 \affiliation{Dipartimento di Fisica e Astronomia, Universit\`a degli Studi di Padova, via Marzolo 8, IT-35131 Padova, Italy}

\author{Ad\'an Cabello}
	\affiliation{Departamento de F\'{\i}sica Aplicada II,
		Universidad de Sevilla,
		41012 Sevilla,
		Spain
	}
	\affiliation{
		Instituto Carlos~I de F\'{\i}sica Te\'orica y Computacional, 
		Universidad de Sevilla, 
		41012 Sevilla, 
		Spain
	}


\begin{abstract}
Entanglement is a key resource in many quantum information tasks. From a fundamental perspective entanglement is at the forefront of major philosophical discussions advancing our understanding of nature. An experimental scheme was proposed in 1989 by Franson that exploited the unpredictability in the generation time of a photon pair in order to produce a then new form of quantum entanglement, known as \textit{energy-time entanglement}. A later modification gave rise to the very popular \textit{time-bin entanglement}, an important cornerstone in many real-world quantum communication applications. Both forms of entanglement have radically pushed forward our understanding of quantum mechanics throughout the 1990s and 2000s. A decade later modifications to the original proposals were proposed and demonstrated, which opens the path for the highly sought-after device-independence capability for entanglement certification, with a goal of ultra-secure quantum communication. In this review we cover the beginnings of energy-time and time-bin entanglement, many key experiments that expanded our understanding of what was achievable in quantum information experiments all the way down to modern demonstrations based on new technological advances. We will then point out to the future discussing the important place that energy-time and time-bin entanglement will have in upcoming quantum networks and novel protocols based on nonlocality.

\end{abstract}
\maketitle



\section{Introduction}


Entanglement is widely considered to be the key feature of quantum mechanics having been referred to by Schr\"{o}dinger himself as ``\textit{the} characteristic trait of quantum mechanics"~\cite{Schrodinger35}. It has represented the most intriguing aspects of nature being at the forefront of historical discussions~\cite{EPR, Bohr, Bell_1964} that have pushed forward our understanding of the quantum domain. More recently, practical applications based on entanglement have emerged. For example, the remarkable connection between entanglement and remote and secure cryptographic key generation was discovered in 1991~\cite{Ekert:1991PRL}, greatly pushing forward the field of quantum communication and information. 

Polarization entanglement has been for decades the workhorse of studies of foundations of quantum mechanics~\cite{Freedman:1972PRL, Aspect:1982PRL, Weihs:1998PRL, Giustina:2015PRL, Shalm:2015PRL} and practical applications of quantum information~\cite{Poppe04, Ono2013, Yin_2020, Schiansky2023}. In spite of its great success as a quantum information resource, the polarization degree-of-freedom has one main limitation which hinders its application in modern quantum information science: the fact that it is limited to encoding of two-dimensional systems (qubits). With the recent discovery of many tasks that are significantly enhanced by high-dimensional entanglement, the use of other degrees-of-freedom that can support high-dimensionality is a necessity~\cite{Erhard2020}.

Franson proposed a novel setup for generating and analyzing a radically different form of entanglement in 1989~\cite{Franson_1989}. It exploited the random nature in the generation time of a photon pair from the cascaded decay in a three-level atomic system, which was then analyzed by unbalanced Mach-Zehnder interferometers \cite{Jin2024}. The scheme was demonstrated already shortly afterwards~\cite{Kwiat1990, Ou1990} and it quickly became the mainstay for long-distance Bell violation experiments during the 90s~\cite{Tapster1994, Tittel1998}. 

At the time, these were the longest distance Bell inequality violations performed, and it showed the potential of quantum communications. In 1999, an important modification to the original Franson setup, which allowed precise time synchronization, was proposed and demonstrated~\cite{Brendel1999}. Originally called pulsed energy-time entanglement, this was later referred to as time-bin entanglement, and it opened up the path for large increases in propagation distance~\cite{Marcikic2004}. 

Apart from Bell tests and entanglement propagation, energy-time and time-bin entanglement were employed as a resource in many applications in quantum information such as entanglement-based quantum key distribution~\cite{Ekert1992} and later its multi-user version~\cite{Wen2022}, quantum random number generation~\cite{Xu2016}, transferring quantum information between photons of different wavelengths~\cite{Tanzilli2005}, used to construct hyper-entanglement~\cite{Barreiro2005}, storage of entanglement onto quantum memories~\cite{Clausen2008}, quantum teleportation~\cite{vanHouwelingen2006} entanglement swapping~\cite{Sun2017}, cluster states for one-way quantum computing~\cite{Reimer2019}, entanglement distillation~\cite{Ecker2021} and quantum secret sharing~\cite{Takesue2006}. 

In this review we will cover both energy-time and time-bin entanglement from their origins, go over many key experiments that have significantly advanced our knowledge of quantum communications and foundations of quantum mechanics, review the current state-of-the-art and give an outlook of the many open challenges as well as point important directions to pursue. 


\section{Energy-time entanglement and early experiments}


In 1964 a now well-known proposal to test the existence of local hidden variables was put forth by John Bell~\cite{Bell_1964}, in which an inequality based on the assumptions of measurement, parameter and outcome independence could be tested in an experiment with correlated particles. Throughout the 70s and 80s experimental violations of Bell inequalities were carried out~\cite{Freedman:1972PRL, Aspect:1982PRL} using polarization entanglement generated from atomic cascades. 

\begin{figure*}[!ht]
 \centering
 \includegraphics[width=18cm]{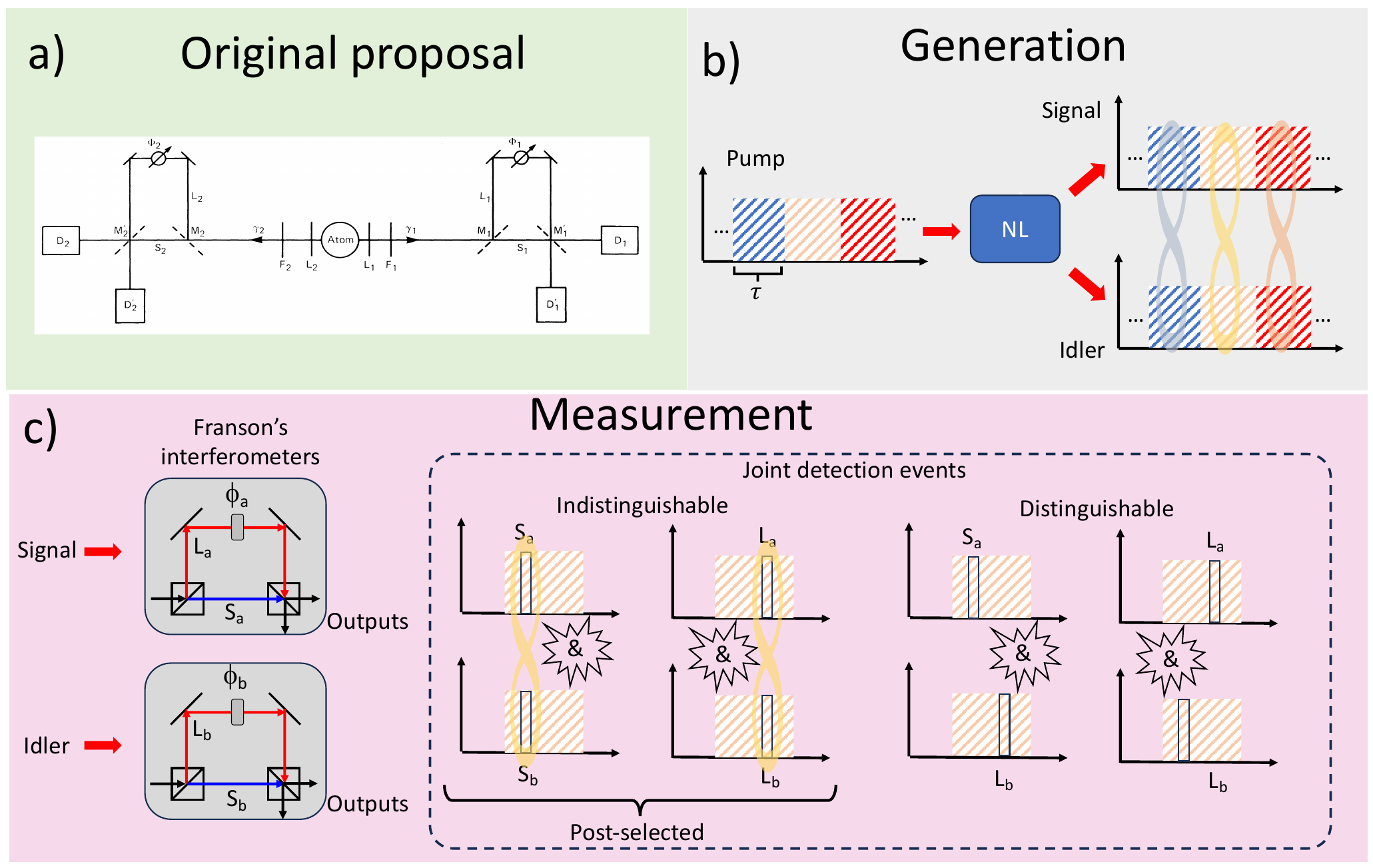}
\caption{Scheme for the generation and measurement of energy-time entanglement. a) Franson's original proposal. Reproduced with permission from ~\cite{Franson_1989} Copyright (1989) American Physical Society. b) A continuous laser beam of coherence time $\tau$ is fed to a non-linear (NL) process, having a small probability of creating a photon pair at a random time within this time~$\tau$. Since the pair is correlated in energy (due to conservation in the NL process) and in creation time, this form of entanglement is referred to as energy-time. c) Unbalanced interferometers, often called Franson's interferometers, are employed which superpose half of the arriving wavefunction joint amplitude probabilities erasing their time information, making these indistinguishable processes entangled. The other half of the coincident detection events can be distinguished in principle, due to the paths taken in the interferometers. These must be post-selected by the detection electronics in order to attain a Bell violation.} 
\label{Fig1}
\end{figure*}

The completely new scheme proposed by Franson in 1989 instead uses entanglement generated from the uncertainty in the generation time of a photon pair (Fig.~\ref{Fig1}a)~\cite{Franson_1989}. It is based on a continuous laser beam, usually referred to as the pump with coherence time $\tau$, which passes through a non-linear (NL) medium having a small probability of generating a photon pair within a time uncertaintly equal to the same $\tau$ (Fig.~\ref{Fig1}b). The two photons, usually referred to as signal and idler are correlated in time and energy, due to the momentum and energy conservation in the non-linear process. 


The NL process produces entangled photons of the form $\int{d\omega_id\omega_s \phi(\omega_i, \omega_s)|\omega_i\rangle|\omega_s\rangle}$, where $\omega_i$ and $\omega_s$ are the frequencies of the generated signal and idler photons, respectively, and $\phi(\omega_i, \omega_s)$ is the amplitude probability of generation of the photon pair at the specific generated photon frequencies, which is dependent on the NL process and pump laser.
In many cases, this two-photon wavefunction
$\phi(\omega_i, \omega_s)$
can be approximated by a double Gaussian function as
\begin{equation}
    \phi(\omega_i, \omega_s)\propto 
    e^{-\frac{\tau^2}{4\pi}(\omega_i+\omega_s-\Omega_p)^2}
e^{-[\delta_i(\omega_i-\Omega_i)-\delta_s(\omega_s-\Omega_s)]^2}\,,
\end{equation} with $\Omega_p$ the pump frequency,
$\Omega_{i,s}$ the central idler/signal frequencies ($\Omega_{i}+\Omega_{s}=\Omega_{p}$)
and $\delta_{i,s}$ related to the phase-matching condition
(see for instance \cite{kolenderski09pra,bennink10pra,jin18prap,coccia23pra}
for the detailed form of $\phi(\omega_i,\omega_s)$ for photon pair sources based on spontaneous parametric down-conversion).
Each photon is then sent to the communicating parties (usually referred to as Alice and Bob) through an appropriate channel. At this point the entanglement is defined over a continuous frequency spectrum. 

Since in many cases the coherence time $\tau$ is much larger than $\delta_{i,s}$, the frequencies of the two photons are mainly anti-correlated, implying
correlation in the detection times. 
Indeed, the two-photon temporal amplitude is calculated as the Fourier
transform of $\phi(\omega_i, \omega_s)$, and it is given by
\begin{equation}
\begin{aligned}
    F(t_i, t_s)=&\int {\rm d}\omega_i{\rm d}\omega_s\,
    \phi(\omega_i, \omega_s) e^{i(\omega_i t_i+\omega_s t_s)}
    \\
    \propto&\,\,
    e^{-\frac{\pi}{\tau^2}(\frac{\delta_i t_i+\delta_s t_s}{\delta_i+\delta_s})^2}
    e^{-(\frac{t_i-t_s}{2(\delta_i+\delta_s)})^2}e^{i(t_i\Omega_i+t_s\Omega_s)}.
\end{aligned}
\end{equation}
The above quantum state
corresponds to the simultaneous emission of two photons;
$F(t_i,t_s)$ is suppressed for $|t_i-t_s|\gtrsim|\delta_i+\delta_s|$, 
with the moment of emission unpredictable within the pump coherence time $\tau$.

The measurement is done using unbalanced interferometers (Fig.~\ref{Fig1}c)~\cite{Franson_1989}, which map the photons to two well defined paths with different lengths, a short ($S_{a, b}$) and a long ($L_{a, b}$) path, where the subscripts $a$ and $b$ correspond to the paths in Alice's or Bob's interferometer respectively. A relative phase ($\phi_a$ or $\phi_b$) can be applied within each interferometer. Finally, the two paths are superposed on a beamsplitter (BS), with the two outputs connected to single-photon detectors. 

The interferometers are tuned such that the length difference between the long and short path follow
\begin{equation}
   \tau \gg (L - S)\gg\tau_p,
\end{equation}
where $\tau_p$ is the coherence time of the photon pair such that no single-photon interference occurs when locally monitoring only Alice or Bob's detectors.
When jointly observing coincidence detections across Alice and Bob, three joint arrival times are possible: early detection at Alice with respect to Bob; coincident detections at Alice and Bob; and late detection at Alice with respect to Bob. These correspond to three subsets of the four different path combinations: $\{S_aL_b\}$,$\{S_aS_b,L_aL_b\}$, and $\{L_aS_b\}$.  The coincident detection events that correspond to equal path length ($S_aS_b$ or $L_aL_b$) are indistinguishable in principle if a second condition is also enforced on Franson's interferometers, that the length difference of the two interferometers are close enough, so that
\begin{equation}
    \bigl|(L_a - S_a)-(L_b - S_b)\bigr|\ll\tau_p.
\end{equation}

The other two events can be distinguished simply because their joint detection timing signature is different ($SL$ vs $LS$). 
After the first beam-splitters of the unbalanced
interferometers, we end up with the final bi-partite state 
 \begin{equation}
\begin{split}
     |\Psi\rangle = \tfrac{1}{2}\bigl(&|S_aS_b\rangle + e^{i(\phi_a + \phi_b)}|L_aL_b\rangle\\ 
     &+ e^{i\phi_b}|S_aL_b\rangle + e^{i\phi_a}|L_aS_b\rangle\bigr).
\end{split}
\end{equation} 

The final beamsplitters in the interferometers transform the arriving modes, into their two outputs as $|S\rangle_{a,b} \rightarrow 1/\sqrt{2}(|0\rangle + i|1\rangle)_{a, b}$ and $|L\rangle_{a, b} \rightarrow 1/\sqrt{2}(|0\rangle - i|1\rangle)_{a, b}$, where $|0\rangle_{a, b}$ and $|1\rangle_{a, b}$ correspond to the final outputs of the beamsplitters at Alice and Bob's intererometers. Applying this transformation to the first two terms of energy-time entangled bi-partite state above, while discarding the other two terms as they are distinguishable (Fig.~\ref{Fig1}c), we obtain the following final state:
\begin{equation}
\begin{aligned}
|\Psi\rangle_{ps} =& \tfrac{1}{2}\bigl[ (|0\rangle + i|1\rangle)_a (|0\rangle + i|1\rangle)_b \\
&+ e^{i(\phi_a + \phi_b)} (|0\rangle - i|1\rangle)_a (|0\rangle - i|1\rangle)_b \bigr]
\\
=& 
\tfrac{1}{2}
\bigl[ \cos\frac{\phi_a+\phi_b}{2}(|00\rangle_{ab} - |11\rangle_{ab}) 
\\
&
+ \sin\frac{\phi_a+\phi_b}{2}(|01\rangle_{ab} + |10\rangle_{ab})   \bigr].
\end{aligned}
\end{equation}

\begin{figure*}[!ht]
 \centering
 \includegraphics[width=18cm]{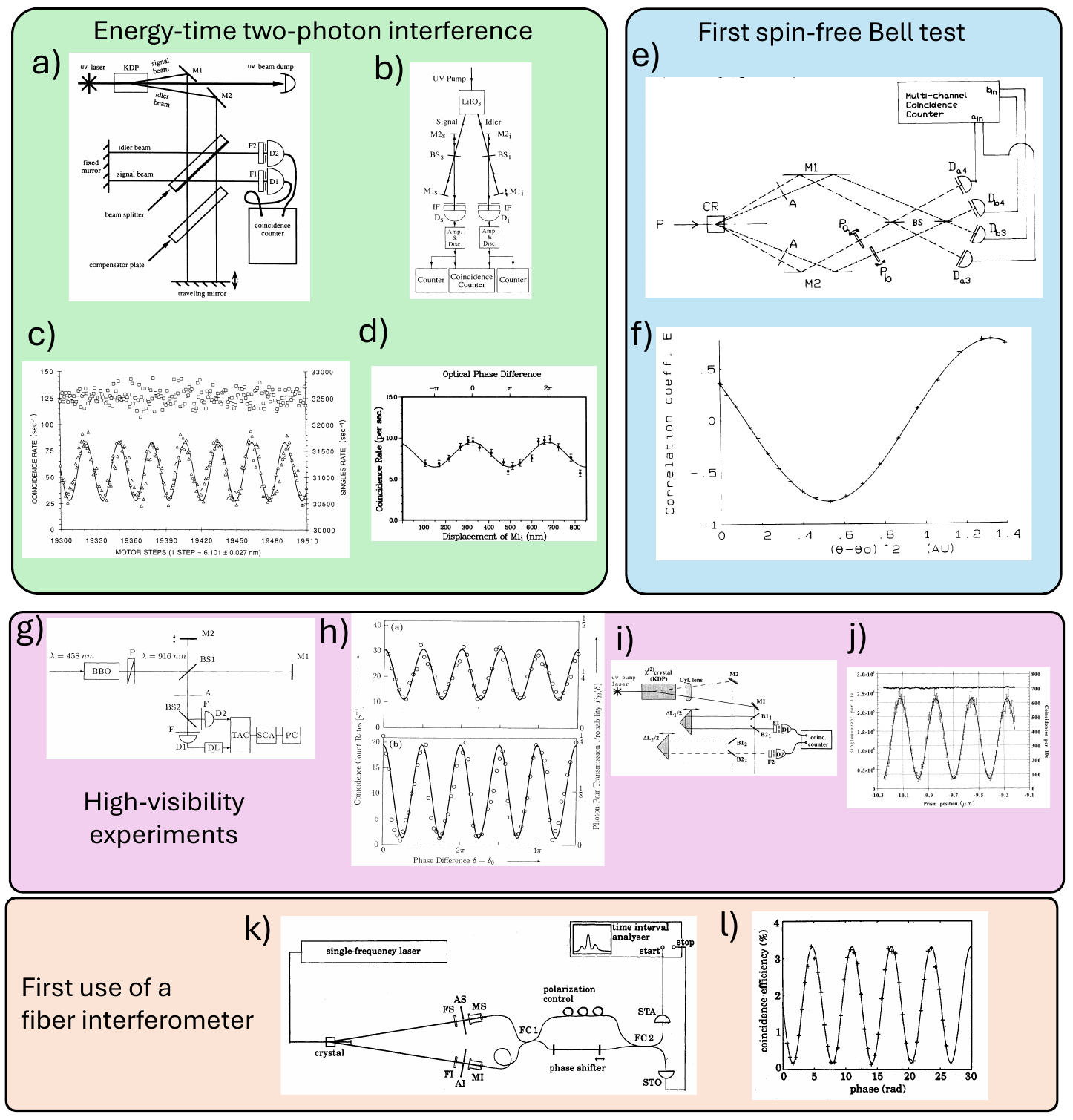}
\caption{Early energy-time experiments. a) and b) show the first implementations of two-photon interference based on energy-time entanglement following Franson's proposal with unbalanced interferometers~\cite{Franson_1989}. Reproduced with permission from a)~\cite{Kwiat1990} and b)~\cite{Ou1990} Copyright (1990) American Physical Society.  c) and d) show the obtained two-photon interference curves for~\cite{Kwiat1990} and~\cite{Ou1990}, respectively, where the two-photon visibility is limited to 50\% since the non-interfering wavefunction components (i.e. short-long) are not removed in these experiments.~\cite{Kwiat1990} also shows the absence of interference in the single-event detections. Reproduced with permission from c)~\cite{Kwiat1990} and d)~\cite{Ou1990} Copyright (1990) American Physical Society. e) First Bell inequality violation that did not rely on the polarization degree of freedom. Reproduced with permission from \cite{Rarity1990a} Copyright (1990) American Physical Society. The experimental setup relies on two-photon interference through two balanced interferometers. f) shows the expectation value for the CHSH Bell inequality. Reproduced with permission from ~\cite{Rarity1990a} Copyright (1990) American Physical Society. g) and h): by introducing a short detection window able to single-out only the overlapping central detection peak corresponding to the short-short and long-long wavefunction components,~\cite{Brendel1991} was first able to demonstrate a higher two-photon visibility than 50\%. Reproduced with permission from ~\cite{Brendel1991} Copyright (1991) American Physical Society. h) specifically shows the difference in visibility when the post-selection procedure is applied (top vs. bottom interference curve). i) and j)~\cite{Kwiat1993} then demonstrated the first two-photon energy-time high-visibility experiment with two separate measurement parties, opening up the path for Bell tests with parties located far apart. Reproduced with permission from ~\cite{Kwiat1993} Copyright (1993) American Physical Society. k) and l)~\cite{Rarity1993} demonstrated the first energy-time two-photon interference experiment with the use of an optical fiber interferometer, and also exploited post-selection to obtain high-visibility. Reproduced with permission from ~\cite{Rarity1993} Copyright (1993) EDP Sciences.} 
\label{Fig2}
\end{figure*}

By projecting the state above onto the outputs at Alice's and Bob's interferometers we obtain joint detection probabilities showing an interference pattern proportional to 
$\cos^2\frac{\phi_a + \phi_b}{2}$
and \
$\sin^2\frac{\phi_a + \phi_b}{2}$
respectively for the same joint outputs or crossed ones. This remarkable non-local two-photon interference effect was demonstrated shortly after Franson's landmark proposal in two independent experiments (Figs.~\ref{Fig2}a-d)~\cite{Kwiat1990, Ou1990}. 

Both experiments employed unbalanced Michelson's interferometers, instead of Mach-Zehnder's as in Franson's proposal, for experimental reasons with no consequence to the expected results. As the non-linear process for photon pair generation, both experiments employed, then in its early stages, spontaneous parametric down-conversion (SPDC)~\cite{Burnham1970}. Very clearly, both show a two-photon interference pattern but with only 50\% visibility, owing to the presence of the non-interfering $|S_aL_b\rangle$ and $|L_aS_b\rangle$ components. Furthermore,~\cite{Kwiat1990} showed the absence of interference when only detecting single detection events (Fig.~\ref{Fig2}c), as expected~\cite{Franson_1989}. At the same time in parallel, a series of other experiments was investigating non-local two-photon interference~\cite{Horne1989} but employing balanced interferometers~\cite{Rarity1990a, Rarity1990b, OuPRA1990}, in which case two-photon visibilities higher than 50\% are reachable, while still having the absence of single-photon interference. Notably~\cite{Rarity1990a} demonstrated using their experimental arrangement the first Bell inequality violation using a degree-of-freedom other than polarization (Figs.~\ref{Fig2}e and \ref{Fig2}f). 

In 1991, Brendel \textit{et al.}\ demonstrated the first two-photon high-visibility experiment using Franson interferometers, by employing a short coincidence window in the detection electronics post-selecting only the overlapping indistinguhishable $|S_aS_b\rangle$ and $|L_aL_b\rangle$ components (Figs.~\ref{Fig2}g and \ref{Fig2}h)~\cite{Brendel1991}. This post-selection procedure does recover a high-visibility Bell inequality violation usable as an entanglement test, but has the downside that it removes the direct link between Bell inequality violations and violations of local realism~\cite{Aerts:1999PRL}; we will return to how to restore this link in Section III. Nonetheless, the benefits of the post-selection procedure caused it to quickly become adopted, which led to the first energy-time Bell inequality violations (Figs.~\ref{Fig2}i and \ref{Fig2}j)~\cite{Brendel1992, Kwiat1993} and also the first use of an optical fiber interferometer two-photon energy-time experiments (Figs.~\ref{Fig2}k and \ref{Fig2}l)~\cite{Rarity1993}, paving the way for what would come later in terms of long-distance Bell inequality violations. It is also worth to notice that unbalanced interferometers have also been extensively employed to measure time-bin states in prepare-and-measure scenarios, such as in \cite{Lubasch2018, Zhang2022}



\section{Long-distance Bell tests and time-bin entanglement}


\subsection*{Long-distance energy-time experiments}


All previous discussed experiments were done in the lab without significant propagation distance for the photon pairs, so there was a need to implement separated measurement parties to ensure measurement independence in a Bell inequality violation~\cite{Aspect:1982PRL}. Energy-time entangled states seemed a very suitable candidate to push forward in this direction, especially since in the preferred transmission channel of optical fibers, this form of entanglement is quite robust to decoherence effects~\cite{Gisin2002}. 

Polarization entanglement, on the other hand, is more susceptible to decoherence coming from residual birefringence fluctuations originating from environmental factors, a phenomenon known as Polarization Mode Dispersion (PMD)~\cite{Hui2023book}. This issue used to be more severe in the past when telecommunication optical fibers had higher PMD coefficients and spontaneous parametric down-conversion sources of entangled photons had very short coherence times. Recent developments have shown successful distribution of polarization entanglement over 248 kms of deployed fiber~\cite{Neumann2022}, but energy-time entanglement remains the more robust entanglement form in optical fiber.

The first experiment focusing on long-distance energy-time two-photon interference was already carried out in 1991 by Franson himself~\cite{Franson1991}, using 51 m of free-space propagation distance between the source and the measurement parties, all placed in the same lab, such that the total propagation for the photons is 102 m. The two-photon visibility was not sufficient to violate a Bell inequality since the coincident detection window width was not narrow enough to only detect the overlapping components of the state. 

\begin{figure*}[!ht]
 \centering
 \includegraphics[width=16cm]{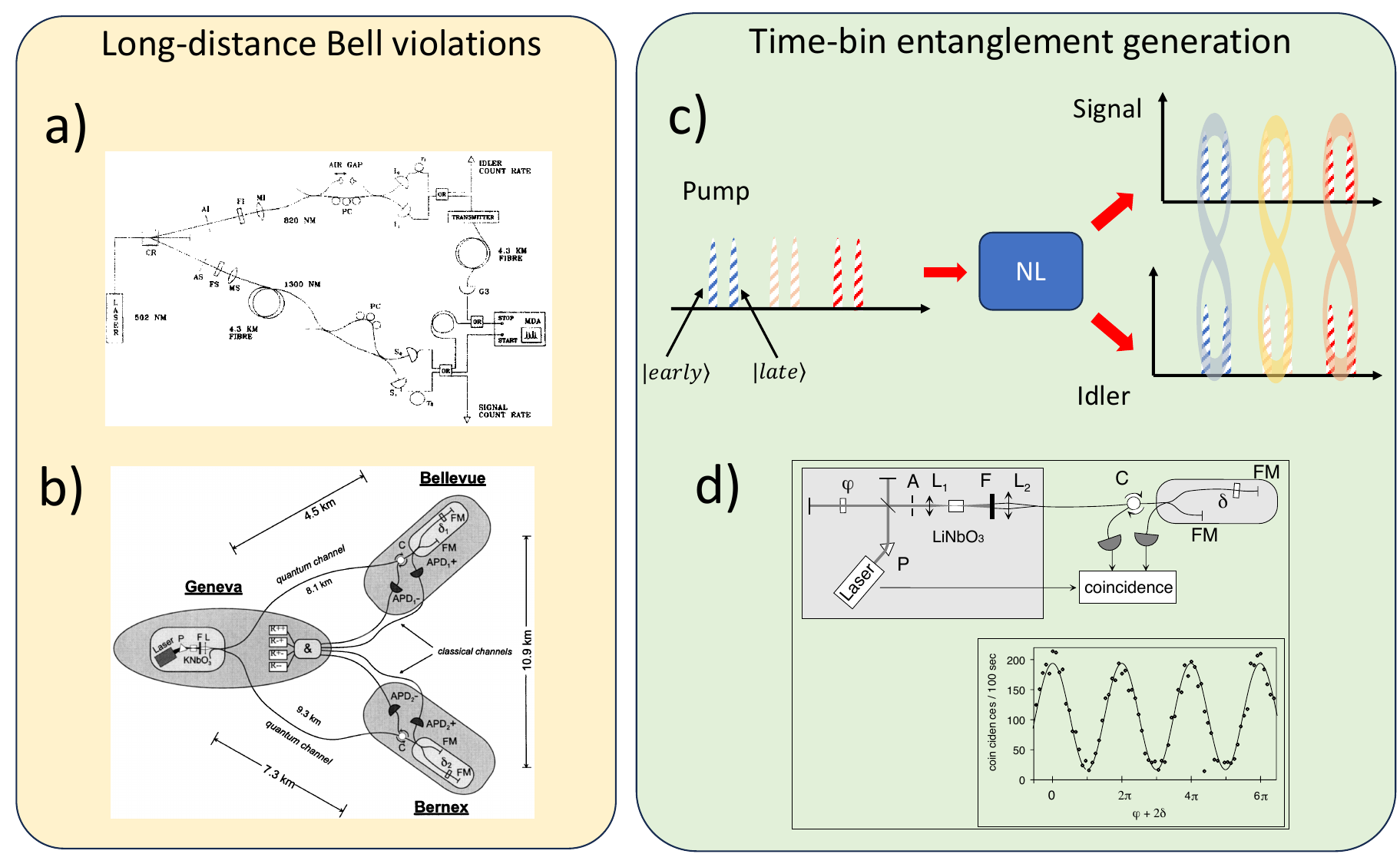}
\caption{Long-distance Bell violations and time-bin entanglement. a) Bell inequality violation using energy-time entanglement with the propagation of one of the photons (at 1300 nm) through 4.3 km of spooled fiber~\cite{Tapster1994}. Reproduced with permission from ~\cite{Tapster1994} Copyright (1994) American Physical Society. b) First energy-time Bell inequality violation over installed optical fibers, with the receiving parties and source connected via optical fibers, and both photons generated at 1300 nm~\cite{Tittel1998}. Reproduced with permission from ~\cite{Tittel1998} Copyright (1998) American Physical Society. c) Scheme to generate time-bin entanglement, where the main difference to energy-time entanglement being that the pump laser is prepared in a coherent superposition of two separate pulses which pass through the non-linear (NL) medium. This causes the photon pairs to only be generated in two well localized ``time-bins'', greatly aiding in synchronization for the detectors, as well as reducing overall background noise. d) First experiment generating time-bin entanglement, using a Michelson interferometer to tailor the pump onto a pulse superposition, and using another Michelson interferometer to analyse the photon pairs showing high-visibility two-photon interference~\cite{Brendel1999}. Reproduced with permission from ~\cite{Brendel1999} Copyright (1999) American Physical Society.} 
\label{Fig3}
\end{figure*}

By taking advantage of optical fibers, fiber-optical interferometers and a narrow coincidence window Tapster \textit{et al.}~\cite{Tapster1994} were able to perform a Bell inequality violation with one of the photons propagating across 4.3 kms through an optical fiber channel, performing the Bell test with the longest propagation distance yet (Fig.~\ref{Fig3}a), showing that entanglement was preserved over much greater distribution distances than previous realizations. This was achieved by tailoring the down-conversion process to generate one of the photons at 1300 nm, having much lower attenuation in optical fibers than the more traditional shorter wavelengths used in previous experiments.


At this point, it was becoming clear that energy-time entanglement was a strong contender to support Bell-type correlations over longer distances. Therefore, different techniques to improve the robustness and efficiency of the experiments were critical to push this forward. The shift to more optimized wavelengths for optical fiber transmission, combined with suitable detectors such as Germanium avalanche photo-diodes were important steps taken, both implemented already by~\cite{Tapster1994}. Interferometers implemented with optical fiber components was another important measure, avoiding additional losses as well as improved robustness when compared to bulk optics systems. Another notable improvement was the use of Michelson fiber-optical interferometers with mirrors combined with $45^{\circ}$ Faraday rotators, the so-called Faraday mirrors (FMs)~\cite{Tittel1997}, auto-compensating birefringence effects following optical fiber transmission. This latest experiment was also the first energy-time Bell inequality violation where the measurement stations were located in different rooms with respect to the source (35 m total physical distance between the parties). 

Then, in 1998, two energy-time experiments were published where both parties were located at several kilometers away from the source connected by optical fibers~\cite{Tittel1998b, Tittel1998}. In both experiments, the source was located in Geneva, while one party was placed in Bellevue and the other in Bernex with 8.1 and 9.3 km of optical fiber propagation, respectively. The actual straight line distances were 4.5 and 7.3 km, respectively, with the separation between both parties being over 10 km (Fig.~\ref{Fig3}b), showing for the first time that quantum correlations held over such distances. This result also strengthened the feasibility of practical entanglement-based quantum communication protocols, such as quantum key distribution. The large technological improvements in order to reach these results, involved tuning the photon pair source to produce both photons at 1310 nm, benefiting from lower attenuation and minimal chromatic dispersion, optimization of germanium single-photon detectors, stabilization of the interferometers and detection electronics.~\cite{Tittel1999}. 

Ten years later, another experiment increased the distance of the optical fiber channel to 100 km, by combining different techniques~\cite{Zhang2008}: (i) the use of planar lightwave circuits (PLC)~\cite{Honjo2004} to implement the unbalanced Franson interferometers and (ii) the use of superconducting single-photon detectors (SSPDs), yielding noticeable improvements in both dark count rate and detection timing jitter~\cite{Goltsman2001, Hadfield2009}. The implementation of the unbalanced interferometers onto planar lightwave circuits (PLC)~\cite{Honjo2004} greatly helped in improving the robustness of the analyzers, since PLCs are highly robust in terms of relative phase stability (like photonic integrated circuits~\cite{Wang2020}), as well as being highly polarization input independent. The use of SSPDs with much lower dark count rates, allowed for a large reduction in accidental count rates, thus supporting successful high-visibility two-photon interference after 100 km of optical fiber. 

Much more recently, distribution of energy-time entangled photons using a periodic-poled lithium niobate (PPLN) waveguide crystal in an all-fiber setup over 150 km was accomplished~\cite{Aktas2016}. Furthermore, the authors exploited the energy correlations from the broadband nature of the spectrum of the emitted photon pairs to show the possibility to distribute the entangled photons to multiple users. Following advances in integrated photonics technology, a demonstration of distribution of energy-time entangled photons produced from an AlGaAs integrated micro-ring resonator through 12 km of installed optical fibers (12 loops of 2 km each) was performed~\cite{Steiner2023}. 


\subsection*{Time-bin entanglement and long-distance Bell inequality tests}


As the distances increase, one aspect becomes more relevant, that is, the fact that no common timing reference exists. Therefore, the detectors need to be continuously active, thus increasing the chances for accidental detection events. This was solved with the proposal of a source of time-bin entanglement~\cite{Brendel1991} where the arrival times of the photon pairs are discretized. 

The principle is that each pump laser photon is already created in a superposition state of the form $\alpha|\textrm{early}\rangle + \beta e^{i\phi_p}|\textrm{late}\rangle$ (Fig.~\ref{Fig3}c), normally with an unbalanced interferometer, where $|\textrm{early}\rangle$ ($|\textrm{late}\rangle$) correspond to an early (late) time-bin, respectively, $\alpha$ and $\beta$ are amplitude coefficients adjusted in the preparation interferometer and $\phi_p$ is a relative phase. The entangled state produced following the NL medium has the form $\alpha|\textrm{early}_i\rangle|\textrm{early}_s\rangle + \beta e^{i\phi_p}|\textrm{late}_i\rangle|\textrm{late}_s\rangle$, where the subscripts $i$ and $s$ correspond to the idler and signal photons, respectively. Another difference of the time-bin entangled photon pair source for practical applications is that there are no strong coherence requirements on the pump laser as opposed to the pump in the energy-time case, since the superposition is prepared in an unbalanced interferometer from a single laser pulse~\cite{Brendel1999}. 

A major advantage concerning the single-photon detection technology available at that time, is the fact that the detectors could be kept inactive (reverse biasing below the breakdown region), and only activated in the time slots when the single-photons could be, as defined by the time-bin locations. This opened up the path for the use of Indium Galium Arsenide (InGaAs) detectors, which are more suitable for long-distance transmission than Germanium, but must operate under gated mode because of too high noise at temperatures reachable by thermoelectric cooling~\cite{Ribordy1998}. Fig.~\ref{Fig3}d shows the first experimental demonstration~\cite{Brendel1999} of a time-bin entanglement source, where a Michelson interferometer is used to prepare the pump superposition from a single optical pulse and a single Michelson interferometer with Faraday mirrors is used as the measurement analyzer. 

\begin{figure*}[!ht]
 \centering
 \includegraphics[width=18cm]{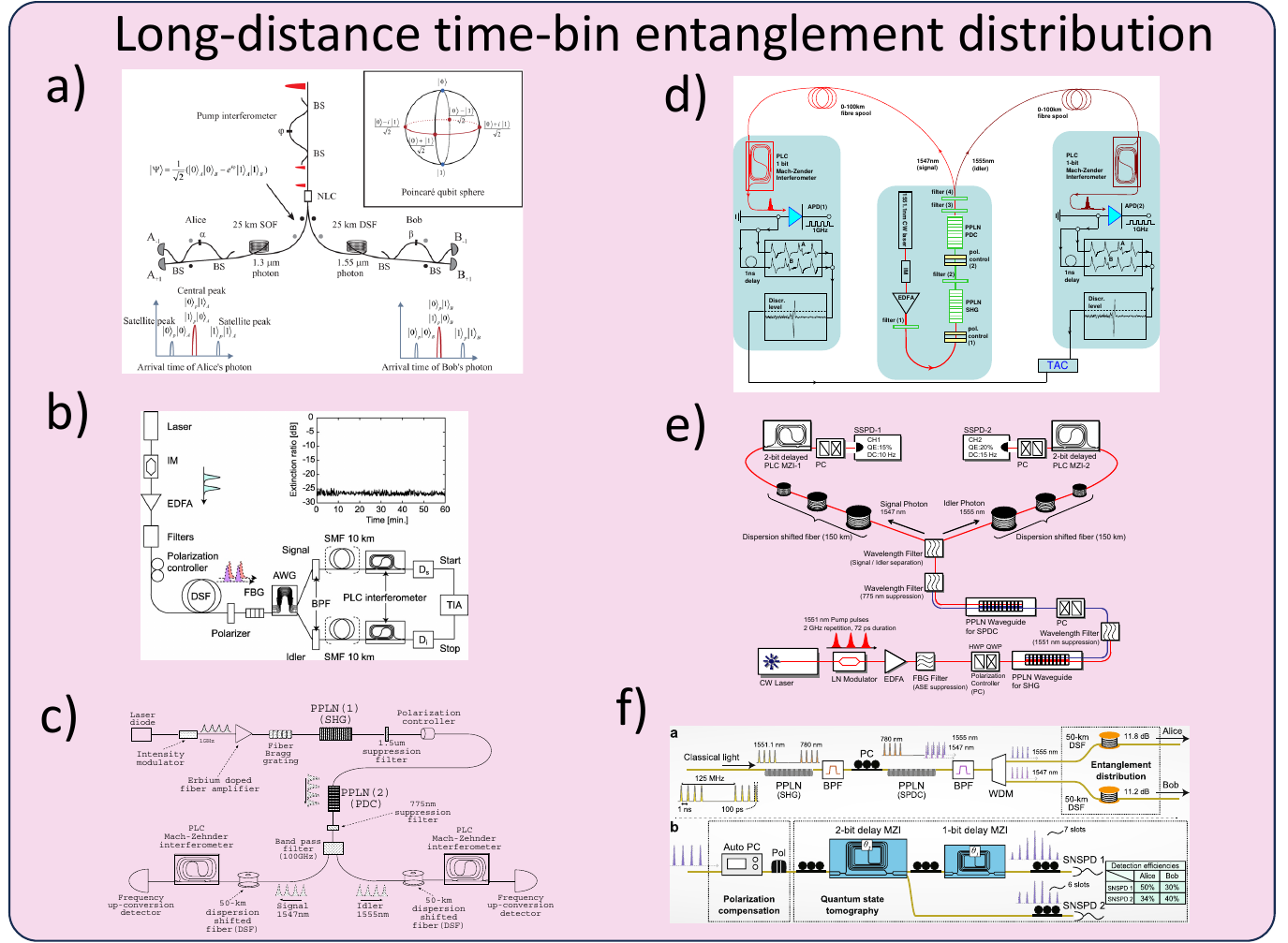}
\caption{Long-distance time-bin entanglement distribution over optical fibers. a) Bell inequality test using non-degenerate photon pairs (1310 nm and 1550 nm), over two fiber spools with a total of 50 km~\cite{Marcikic2004}. Reproduced with permission from \cite{Marcikic2004} Copyright (2004) American Physical Society. b) Use of a long coherence pump laser, dispensing the use of a pump interferometer combined with the use of stable and compact planar lightwave circuits (PLCs) to implement Franson's interferometers for the analyzers, thus demonstrating a Bell violation over 20 km of total spooled fiber~\cite{Takesue_2005}. Reproduced with permission from \cite{Takesue_2005} Copyright (2005) American Physical Society. c) Experiment with a similar approach but replacing avalanche gated-mode single-photon detectors with up-conversion technology supporting a much higher repetition rate enabling a successful Bell violation over a total of 100 km~\cite{Honjo2007}. Reprinted with permission from \cite{Honjo2007} \copyright (2007) Optical Society of America. d) With the use of self-diferencing avalanche single-photon detectors~\cite{Dynes2009} was able to push the transmission distance much further to a total of 200 kms. Reprinted with permission from \cite{Dynes2009} \copyright (2009) Optical Society of America. e) Bell inequality violation over 300 km taking advantage of superconducting single-photon detectors~\cite{Inagaki2013}. Reprinted with permission from \cite{Inagaki2013} \copyright (2013) Optical Society of America. f) Propagation of four-dimensional time-bin entanglement over 100 km~\cite{Ikuta2018}. Reprinted from \cite{Ikuta2018} Copyright (2018) Authors under a CC BY 4.0 license.} 
\label{Fig4}
\end{figure*}

These advances combined allowed Bell inequality tests and other applications over much longer distances than previously possible. A first experiment already studied the preservation of maximal and partial time-bin entanglement over 11 km~\cite{Thew2002}. Then, in 2004, Marcikic \textit{et al.}\ demonstrated the successful distribution of time-bin entanglement over 50 km of spooled optical fiber in the lab~\cite{Marcikic2004}, marking the longest distance yet over which entanglement could be successfully measured. This paper exploited non-degenerate photons at 1310 and 1550 nm, propagating through 25 km long fiber spools of standard and dispersion-shifted (DS) single-mode fiber, respectively (Fig.~\ref{Fig4}a). The use of DS fiber for the 1550 nm photon at such long distances is critical to ensure a zero chromatic dispersion coefficient and thus ensure the different arrival times are resolvable by the unbalanced interferometer. 

Other groups then focused on different technological approaches. One such demonstration was the long-distance propagation of time-bin entanglement over a total distance of 20 km of spooled fibers~\cite{Takesue_2005} (Fig.~\ref{Fig4}b). This experiment was notable for the generation of photon pairs based on the non-linear process of spontaneous four wave mixing (SFWM) within a dispersion-shifted optical fiber~\cite{Fiorentino2012, Li2005}. This is a non-linear $\chi^{(3)}$ process, instead of the $\chi^{(2)}$ nature of the more popular spontaneous parametric down-conversion. Therefore SFWM requires two pump photons in order to generate a signal and idler correlated pair, with these photons being necessarily non-degenerate centered around the pump. An advantage is that the entire interaction occurs already within the same spatial mode in a single-mode optical fiber, thus avoiding losses when coupling from a bulk non-linear SPDC crystal to a fiber. Disadvantages include the fact that the $\chi^{(3)}$ process is less efficient, thus requiring higher optical pump power (typically pulsed light), and additional noise generated from spontaneous Raman scattering within the optical fiber medium comprising the source. 

Due to the nature of the SFWM process, the pump photons are within the telecom spectral window, and long coherence lasers are widely available at these wavelengths, there is no need to employ a pump interferometer to create the time-bin superposition. This can be directly prepared with electro-optical amplitude and phase modulators, and then the pump photons are directly fed to the SFWM optical fiber~\cite{Takesue_2005}. Following this first experiment and employing the same techniques, including PLC circuits for the Franson interferometers, a total propagation distance of 60 km was then demonstrated relying on two 30 km spools of DS fiber~\cite{Takesue2006b}. 

Subsequent experiments continued the trend of increasing the propagation distance for time-bin entanglement by focusing on the more efficient SPDC process and combining with a long-coherence pump laser dispensing the pump interferometer, as well as exploiting different single-photon detection technologies~\cite{Honjo2007, Dynes2009}. However, compared to SWFM, pumping must be done at shorter wavelengths as only one pump photon is involved, and thus the 1550 nm seed laser for the pump is first frequency-doubled to near infrared wavelengths through a second harmonic generation (SHG) process. In the first of these experiments frequency up-conversion detectors are employed in order to bypass the limitations imposed by the relatively low repetition rate of gated avalanche photodiode single-photon detectors~\cite{Honjo2007}. 

The principle behind up-conversion single-photon detection is to up-convert 1550 nm single-photons to visible or near infrared wavelengths, and then detect them using free-running high-efficiency silicon single-photon detectors~\cite{Albota2004}. Due to the non-gated operation as well as lack of afterpulsing from silicon detectors, the authors in~\cite{Honjo2007} were able to push a repetition rate of 1 GHz. Self-differencing InGaAs avalanche detectors~\cite{Yuan2007}, which are designed to greatly minimize afterpulse noise and thus can operate at much higher rates than standard gated-mode detectors, were employed in~\cite{Dynes2009} to demonstrate a Bell inequality violation after the propagation of time-bin entangled states after a total of 200 km of DS spooled fiber (Fig.~\ref{Fig4}d). Then another experiment replaced the self-differencing APDs with superconducting detectors to demonstrate a successful Bell violation after 300 km of spooled DS fiber~\cite{Inagaki2013} (Fig.~\ref{Fig4}e). This is still the current record in terms of a quantum communication protocol using entangled states over optical fibers. 

Another successful experiment employing the techniques just discussed was the long-distance distribution of four-dimensional time-bin entangled states (characterized with quantum state tomography) over 100 km of DS spooled fibers~\cite{Ikuta2018} (Fig.~\ref{Fig4}f). Here, the use of a long-coherence telecom pump laser being frequency-doubled is critical in order to avoid a complex multi-arm pump interferometer. The detection is based on the use of two unbalanced interferometers placed in series and with delays of one and two pulse separation times, respectively~\cite{Islam2017}. 

Very recently, distribution of fully controllable entangled time-bin qubits over 100 kms of optical fibers was carried out, with the adjustment performed through the amplitude and phase modulators used to tailor the time-bin state encoding on the telecom pump laser~\cite{Kim2024}.


\section{Genuine energy-time and time-bin entanglement}


A Bell test's main goal is to rule out the existence of hidden variables governing the behaviour of a quantum system~\cite{Bell_1964}. The test consists of an inequality whose terms are the result of joint measurements carried out on the subsystems of entangled states. The main assumption in deriving the inequality is that measurement setting choices at one site cannot instantly affect measurement outcomes at another site. Surprisingly, entangled systems violate the inequality, meaning that if there are hidden-variables governing the behaviour of the particles, the model must then be non-local. 

A critical aspect is that hidden-variable models can reproduce the results of the violation by exploiting experimental limitations of implementations of Bell inequality tests~\cite{Larsson:2014JPA}. The most well-known limitations are the locality~\cite{Bell_1964} and the detection~\cite{Pearle:1970PRD} loopholes. The first one is related to the fact that the measurement settings have to be causally independent and although it was already considered in Aspect's seminal experiment~\cite{Aspect:1982PRL}, it was only fully closed in 1998 with the measurement settings dynamically chosen with quantum random number generators and with the measurement parties located 400 m apart~\cite{Weihs:1998PRL}. Both of these experiments employed polarization entanglement.

The detection loophole allows a classical explaination for the violation taking advantage of the fact that not all generated photon pairs are measured due to losses in the generation, transmission and detection of the single-photons. The value for the minimum needed joint detection probability for the maximum violation of the CHSH Bell inequality is 83$\%$~\cite{Garg1987PRD,Larsson1998}, and this value can be lowered depending on the measurements employed and the prepared entangled state itself~\cite{Eberhard1993PRA,Larsson2001}. 

This threshold was unattainable for many years because the efficiency of single-photon detectors was not high enough to produce the required joint efficiencies even without any transmission distance. In fact the first Bell violation experiment to close the detection loophole was performed on trapped ions~\cite{Rowe:2001NAT}, which were located together in the same trap, and thus could not support measurement independence. A breakthrough came much later based on the use of ultra-efficient superconducting transition edge sensors (TES)~\cite{Giustina:2013NAT, Christensen2013}, allowing the closure of the detection loophole in photonic Bell tests for the first time, relying on polarization entangled photon pairs. 

Shortly afterwards, the so-called loophole-free tests, as they closed the locality and detection loophole simultaneously were performed~\cite{Giustina:2015PRL, Shalm:2015PRL, Hensen_2015}, with the first two employing photonic polarization entanglement and superconducting detectors. Recent proposals are gaining track to simultaneously close the locality and detection loopholes by randomly routing the location of the measurement hardware \cite{Chaturvedi2024, Lobo2024, roydeloison2024}. Remarkably, no experiments have so far closed either the detection or locality loophole with energy-time or time-bin entanglement.

The closure of loopholes is not only important from a fundamental understanding of nature, but it has consequences for applications in information security in the device-independent (DI) paradigm~\cite{Larsson2002a,Acin:2007PRL}. There is furthermore one additional major loophole that rises from the temporal post-selection procedure needed in Franson interferometry~\cite{Aerts:1999PRL}. 

This problem was already recognized very early on in the development of energy-time entanglement as clearly expressed in~\cite{Strekalov1996PRA}: ``However, the experiments relying on it [temporal post-selection] are still valid for testing Bell-type inequalities \textit{if an additional assumption is made that the photons in the subensemble of discarded events are not any different from those we choose to look at} (emphasis added).'' 

This paper also reported the first experiment to remove the need for post-selection from energy-time Bell tests. It achieved this through a modification of Franson's interferometers changing the standard 50:50 beamsplitters with polarization beamsplitters (PBS), and using a source of polarization entangled photons. In this way, the PBSs route the photons through the unbalanced interferometers such that the polarization correlation ensures that no combination of short-long paths are possible~\cite{Strekalov1996PRA}. Finally, polarizers placed before the single-photon detectors erase the which-path information, which exists if the polarization information can be retrieved. With this setup 95\% two-photon interference was achieved, without the need for post-selection. A more recent experiment employed this concept over a 1.2 km free-space link~\cite{Steinlechner2017}.

\begin{figure*}
 \centering
 \includegraphics[width=1.0\linewidth]{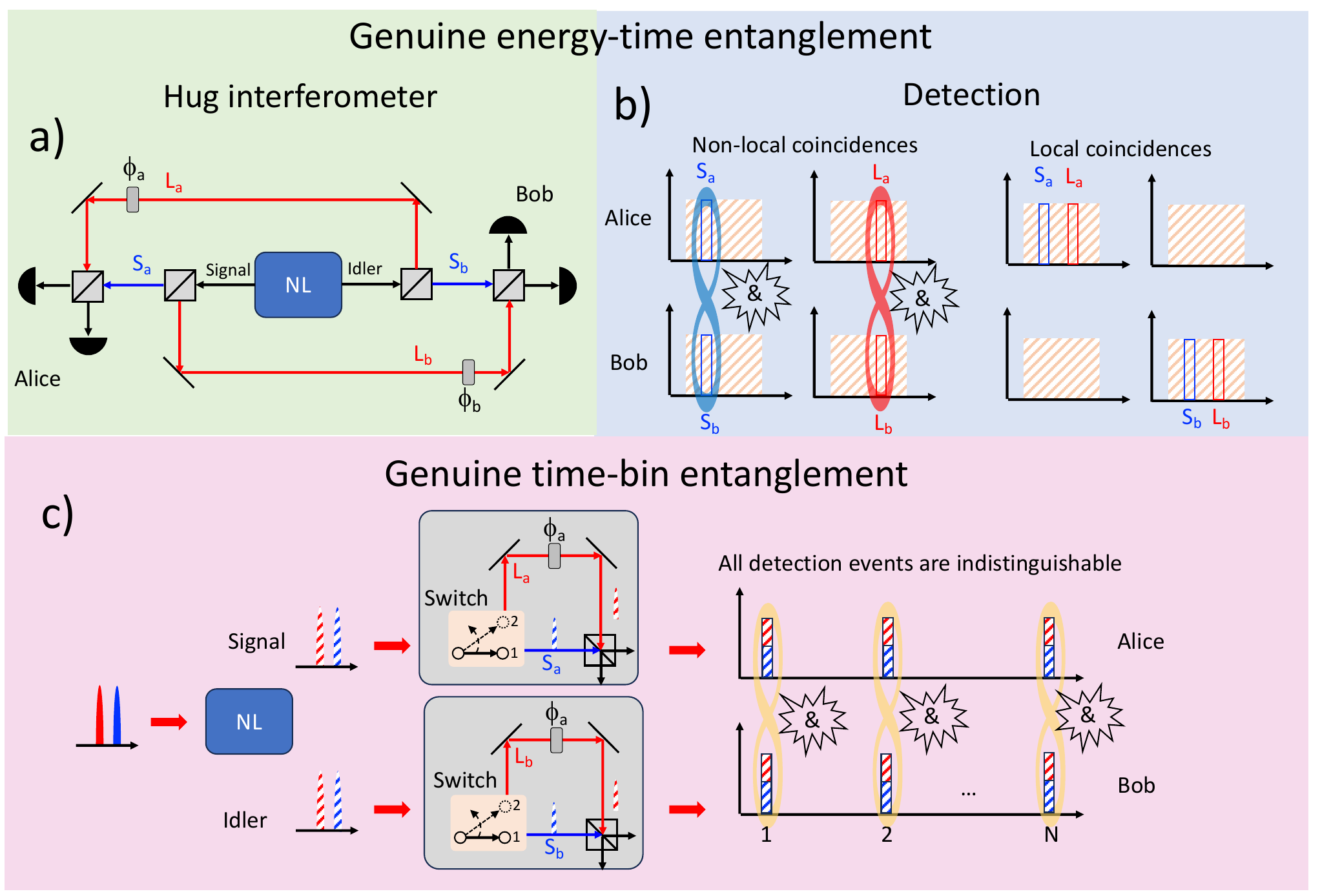}
 \caption{Schemes for genuine energy-time and time-bin entanglement. a) The hug interferometric scheme, designed to remove the post-selection loophole from energy-time entanglement~\cite{cabello_proposed_2009}. The main difference compared to the traditional Franson configuration~\cite{Franson_1989} is that the long arms are routed to the opposing party, thus requiring two parallel communication channels between the source and each communicating party. b) Due to the hug configuration only coincident events where both photons took the same kind of path (short or long) can be detected as non-local events within the coherence time of the pump laser (indicated in yellow). On the other hand, events where the photons from the same pair take different kinds of path are only jointly detected locally. c) In the case of genuine time-bin entanglement the source is unmodified. The analysers on the other hand have the input beamsplitters replaced by optical switches synchronized with the source such that the early generated photons always pass through the long path and the opposite with the late generated photons, such that all amplitude probabilities are always superposed, and no post-selection is needed.}
 \label{fig5}
\end{figure*}

Another direction focuses on ``genuine'' energy-time entanglement in the sense that it does not require another degree-of-freedom. The first approach rearranges Franson's interferometers in order to eliminate non-local non-interfering events~\cite{Cabello2009PRL}. This scheme reconfigures the long arms such that they are rerouted to the opposing party (Fig.~\ref{fig5}a), based on earlier ideas to recombine the paths to remove non-overlapping events~\cite{rossi_generation_2008}. Because of this crossed configuration, this scheme was later named the ``hug'' ~\cite{Cuevas2013NC}. 

Due to this rearrangement, all non-local coincidences occur together within the emission time of the pair, since in this case they must originate only from $|S_aS_b\rangle$ and the $|L_aL_b\rangle$ components. The distinguishable components ($|S_aL_b\rangle$ and $|L_aS_b\rangle$) always arrive to the same party, and thus are post-selected locally (Fig.~\ref{fig5}b). This rearrangement directly removes the need for any post-selection which requires communication between Alice and Bob, and thus no local hidden-variable model can give rise to a loophole based on post-selection. 

In the case of time-bin entanglement a different approach was already suggested in the original proposal for a source of time-bin entangled photons~\cite{Brendel1999}. This approach uses an optical switch synchronized with the source, in order to deterministically route the short/long generated photons to the long/short respective arms of the unbalanced Mach-Zehnder interferometer, thus causing all path amplitude probabilities to overlap simultaneously removing the need for post-selection (Fig.~\ref{fig5}c). 


\begin{figure*}
 \centering
 \includegraphics[width=1.0\linewidth]{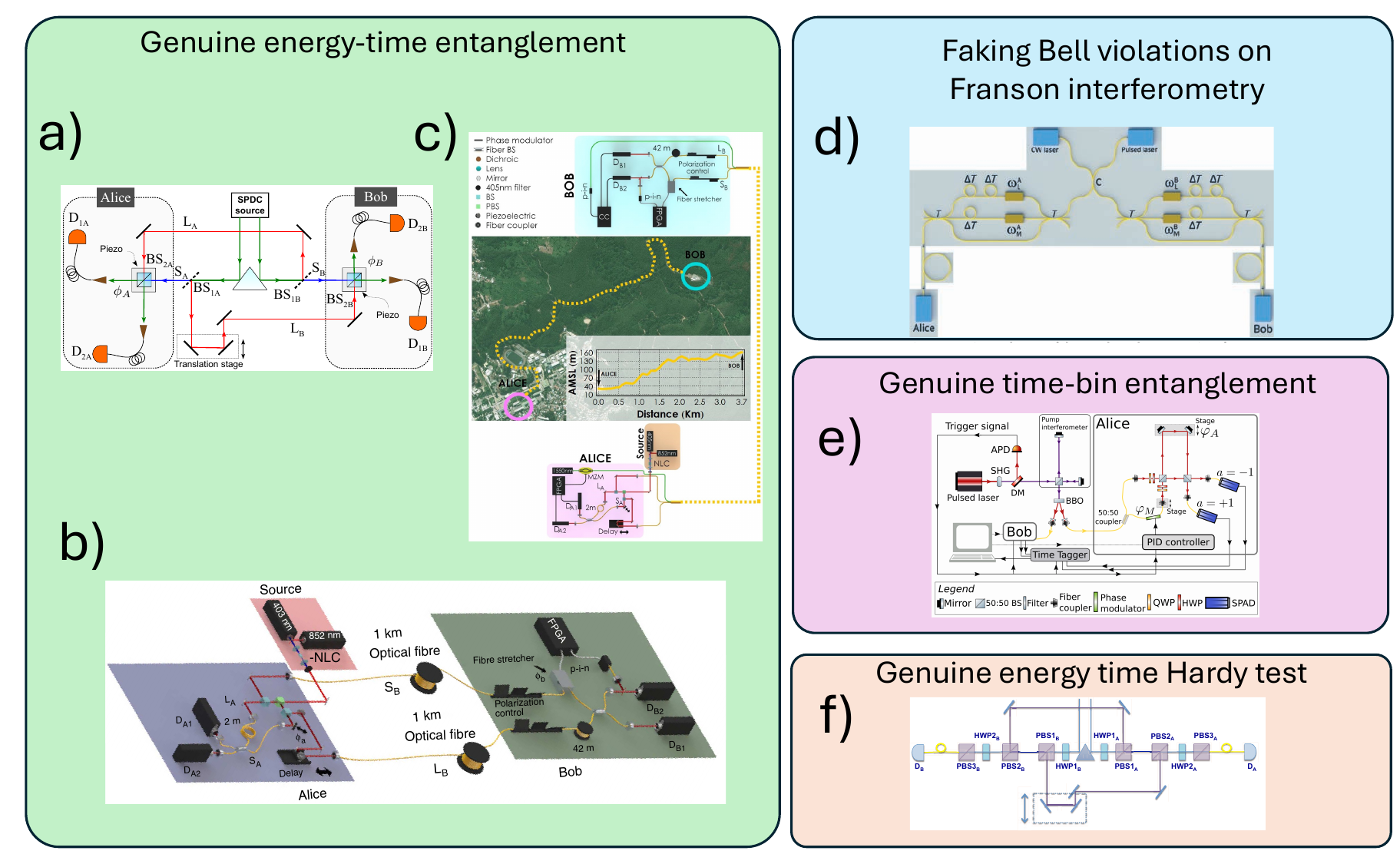}
 \caption{Experiments tackling the post-selection loophole. a) First genuine energy-time Bell test using the hug configuration~\cite{Lima2010PRA}. Reproduced with permission from \cite{Lima2010PRA} Copyright (2010) American Physical Society. b) Genuine energy-time Bell test with the hug where one of the photons propagated over 1 km of spooled optical fibers. Active phase stabilization is employed to compensate the fast phase drift acting on the 1 km long interferometer~\cite{Cuevas2013NC}. Reprinted from \cite{Cuevas2013NC} Copyright (2013) Authors under a CC BY-NC-ND 3.0 license. c) Hug configuration deployed over a 3.7 km deployed optical fiber link showing a successful Bell inequality violation with genuine energy-time entanglement~\cite{CarvachoPRL2015}. Reproduced with permission from \cite{CarvachoPRL2015} Copyright (2015) American Physical Society. d) Exploiting a standard energy-time entanglement setup based on Franson's interferometers by controlling the single-photon detectors with tailored bright light in order to fake the violation results~\cite{Jogenfors2015SA}. Reprinted with permission from \cite{Jogenfors2015SA} Copyright (2015) American Association for the Advancement of Science. e) First genuine time-bin Bell test where active fast switches built from Mach-Zehnder interferometers replace the input beamsplitters in the Franson interferometers, thus removing the post-selection procedure~\cite{Vedovato2018PRL}. Reproduced with permission from \cite{Vedovato2018PRL} Copyright (2018) American Physical Society. f) Demonstration of the hug configuration for genuine energy-time entanglement for other fundamental applications, such as demonstrating a Hardy test free of the post-selection loophole~\cite{Vallone2011}. Reproduced with permission from \cite{Vallone2011} Copyright (2011) American Physical Society.}
 \label{fig6}
\end{figure*}

A violation of a Bell inequality based on the hug interferometric scheme was implemented shortly after the proposal in a table top experiment~\cite{Lima2010PRA}, performing the first non-locality test without the need for post-selection originating from genuine energy-time entanglement (Fig.~\ref{fig6}a). A difficulty of the hug interferometer is the requirement of relative phase stability within the double path channels connecting the source to the parties~\cite{cabello_proposed_2009}. 

In parallel, a practical technological solution to stabilize long-distance fiber-optical interferometers was being developed for general applications in quantum communications~\cite{Cho2009, Xavier2011}. It relies on the use of fiber-optical stretchers based on piezo-electric actuators and real-time feedback using another wavelength, and active compensation over many kilometers long was demonstrated. Based on this active stabilisation technology a first Bell inequality violation of genuine energy-time entanglement was performed over 1 km distance connecting the source to Bob and Alice placed by the source~\cite{Cuevas2013NC} (Fig.~\ref{fig6}b). This work showed the practicality of the hug configuration for removing the post-selection loophole over long distances without the need for other forms of entanglement, and thus could be an important tool for the future of device-independent quantum communication relying on energy-time entanglement. 

The same experimental technique for phase stabilisation was then used in another genuine energy-time Bell test with Bob now separated from the source over 3.7 km of installed optical fibers~\cite{CarvachoPRL2015} (Fig.~\ref{fig6}c), showing the robustness of the hug configuration with active phase stabilization even through deployed optical fibers under real conditions. One may argue the price to close the post-selection loophole may be steep, as one can assume that the post-selected events will not be exploited by nature itself to fake the violation results. However, in practical applications the loophole can be actively exploited in detriment of the users of the system. 

Especially in quantum key distribution, a malicious eavesdropper may take advantage of the presence of the loophole to hack the system, and effectively create a key that Alice and Bob will share. Such an attack was specifically demonstrated in the case of an entanglement-based QKD system based on energy-time entanglement with the Franson configuration~\cite{Jogenfors2015SA} (Fig.~\ref{fig6}c). The authors relied on the technique of sending classical light to blind the single-photon detectors and control them~\cite{Lydersen2010}, showing the Bell test can be faked and even the violation can be steered to a value higher than allowed by quantum mechanics.

The removal of the post-selection loophole in the case of time-bin entanglement has only been performed more recently with the use of active switches as proposed in~\cite{Brendel1999} in the implementation of~\cite{Vedovato2018PRL} (Fig.~\ref{fig6}e). Each ultra-fast switch is implemented with a Mach-Zehnder interferometer containing a phase modulator with nanosecond-response time in one of its arms. 

Very recently, the hug interferometer has been demonstrated as being capable of measuring time-bin entanglement with only local post-selection, thus not being affected by the loophole. This experiment was performed with the hug implemented on a silicon nitride photonic integrated circuit (PIC), showing that it can be used as a compact tool for certification of genuine energy-time and time-bin entanglement~\cite{Santagiustina2024}. Therefore, the hug is actually a general measurement scheme for both genuine energy-time and time-bin entanglement and becomes a crucial tool for future device-independent quantum communication schemes based on these forms of entanglement. 

Finally, the hug has also been demonstrated for other fundamental tests in quantum information such as a Hardy test free of the post-selection loophole~\cite{Vallone2011}. Going forward although the requirement of active phase stabilization may seem highly cumbersome, recent developments in the area of twin-field quantum key distribution have demonstrated active phase stabilization in optical fiber interferometers with independent phase-locked laser sources over hundreds of kilometers~\cite{Wang2022}, thus showing a potential to deploying the hug interferometer in a wide scale. 


\begin{table*}
 \centering
 \begin{tabular}{cccccccccc}
 \hline
 Entanglement type & Distance & Rate & Visibility & Source type & Detectors & Post-selection & Accidental subt. & Year & Reference \\
 \hline \\
 Energy-time & 105 m & 0.005 Hz & $27 \pm 4\%$ & SPDC & ** & No & Yes & 1991 &~\cite{Franson1991} \\
 Energy-time & 4 km & 1 kHz & $86.9 \pm 1.2\%$ & SPDC & Si/Ge APDs & Yes & Yes & 1994 &~\cite{Tapster1994}\\
 Energy-time & 35 m (field) & 75 Hz & $95.7 \pm 3.1\%$ & SPDC & Si/Ge APDs & Yes & Yes & 1997 &~\cite{Tittel1997}\\
 Energy-time & 10.9 km (field) & 25 Hz & $81.6 \pm 1.1\%$ & SPDC & Si/Ge APDs & Yes & Yes & 1998 &~\cite{Tittel1998b} \\
 Energy-time & 10.9 km (field) & 6.5 Hz & $85.2\%$ & SPDC & Si/Ge APDs & Yes & No & 1998 &~\cite{Tittel1998} \\
 Time-bin & 11 km & 3.3 Hz & $86.8\%$ & SPDC & Ge APDs & Yes & No & 2002 &~\cite{Thew2002}\\
 Time-bin & 50 km & 5.5 Hz & $78.0 \pm 1.6\%$ & SPDC & Ge/InGaAs APDs & Yes & No & 2004 &~\cite{Marcikic2004}\\
 Time-bin & 20 km & 2.25 Hz & $99\%$ & SPDC & InGaAs APDs & Yes & Yes & 2005 &~\cite{Takesue_2005} \\
 Time-bin & 60 km & 0.3 Hz & $75.8\%$ & SFWM & InGaAs APDs & Yes & No & 2006 &~\cite{Takesue2006b}\\
 Time-bin & 100 km & 1.5 Hz & $81.6\%$ & SPDC & Up-conversion & Yes & No & 2007 &~\cite{Honjo2007} \\
 Time-bin & 200 km & 0.25 Hz & $79.7\%$ & SPDC & InGaAs SPDa & Yes & No & 2009 &~\cite{Dynes2009} \\
 Time-bin & 300 km & 0.03 Hz & $86.1 \pm 6.8\%$ & SPDC & SNSPDs & Yes & No & 2013 &~\cite{Inagaki2013} \\
 Energy-time & 1 km & 100 Hz & $84.36 \pm 0.47\%$ & SPDC & Si APDs & Not needed & No & 2013 &~\cite{Cuevas2013NC} \\
 Energy-time & 3.7 km (field) & 40 Hz & $82.10 \pm 3.87\%$ & SPDC & Si APDs & Not needed & No & 2015 &~\cite{CarvachoPRL2015} \\
 Energy-time & 150 km & 9 Hz & $89.4 \pm 3.5\%$ & SPDC & InGaAs APDs & Yes & No & 2016 &~\cite{Aktas2016} \\
 Time-bin 4D & 100 km & ** & $0.935 \pm 0.015^{\dagger}$ & SPDC & SNSPDs & Yes & ** & 2018 &~\cite{Ikuta2018} \\
 Time-bin & 100 km & 11.3 kHz & $94.75 \pm 0.16\%$ & SPDC & SNSPDs & Yes & No & 2024 &~\cite{Kim2024} \\
 Time-bin HE & 50 km & 9.8 bps & ** & SPDC & InGaAs SPDs & Yes & ** & 2024 &~\cite{Zhong2024} \\
 \end{tabular}
 \caption{Key parameters of energy-time and time-bin entanglement experiments focused on long-distance propagation. ``**'' Not explicitly given. $^{\dagger}$ State fidelity. APDs: Avalanche Photodiode single-photon detectors; HE: Hyper-entanglement; SFWM: Spontaneous Four Wave Mixing; SNSPDs: Superconducting Nanowire Single-Photon Detectors; SPDC: Spontaneous Parametric Down-Conversion.}
 \label{tab:my_label}
\end{table*}


\section{Integrated sources of energy-time and time-bin entanglement}


Spontaneous parametric down-conversion (SPDC) has been the workhorse behind the generation of earlier energy-time and time-bin entangled photon pairs~\cite{Kwiat1990, Brendel1991, Tapster1994, Tittel1998, Brendel1999, Marcikic2004}. In recent years, other techniques have been developed with the aim of building more compact, integrated and higher efficiency photon pair sources. 


\subsection*{Waveguides}


An early development to improve the generation efficiency of SPDC is the combination of a guided waveguide structure within a non-linear crystal together with a quasi-phase matching (QPM) technique~\cite{Tanzilli_2001}. The use of a waveguide greatly increases the interaction length for the SPDC process, while the QPM, which consists of a periodic inversion of the sign of the $\chi^{(2)}$ crystal nonlinearity generated from a poling process, allows much more flexibility in optimizing the phase-matching for the wavelengths of the pump, signal and idler photons when compared with birefringence phase matching. Another advantage of the use of a waveguide is the possibility to couple input and output optical fibers directly to the crystal, greatly helping with integration and robustness~\cite{Kaiser_2016}.

The use of a waveguide structure within a periodically poled bulk crystal for the efficient generation of energy-time and time-bin entanglement has since then been used in many demonstrations~\cite{Tanzilli_2002, Takesue2005b, Honjo2007_2, Ma2009, Zhong2012, Xie2015, Cheng2023, Mueller2024, Zhang2008, Dynes2009, Inagaki2013, Chang2023, Chang2024b}. More recently new advances in thin-film lithium niobate technology, where the waveguide is considerably shallower, allowed considerable gains in generation efficiency, as well as ultra-broad band emission~\cite{Zhao_2020, Javid2021, Xue2021, Huang2022} (Fig.~\ref{fig7}a). One demonstrated technique that is relevant for ultra-broad band emission is the measurement of geometric phases due to their non-dispersive nature \cite{Jha2018}.

Another move towards integration is the use of cascaded processes within a single non-linear crystal~\cite{Hunault2010}, where SHG and SPDC processes are combined in the same crystal to employ pump lasers at telecom wavelengths. This technique is a simplification compared to many experiments that employed two separate crystals for the SHG and SPDC processes~\cite{Honjo2007, Zhang2008, Dynes2009, Inagaki2013, Ikuta2018, Ma2009}. Cascaded non-linearities within a single crystal continue to be explored recently for energy-time and time-bin entanglement generation~\cite{Lefebvre2021, Zhang2021}. 

Semiconductor materials, specially AlGaAs, have also been explored to build sources of energy-time and time-bin entangled photon pairs, majorly motivated from the possibility of integration with other components, such as pump lasers and detectors which is more natural then with the usual non-linear crystals such as lithium niobate. First attempts aimed at the generation of polarization entanglement from an AlGaAs waveguide~\cite{Orieux2013}, and then also showed two-photon time correlation from an SPDC process within the structure of a diode laser, effectively being a source of photon pairs without requiring an external optical pump~\cite{Boitier2014}. The other experimental demonstrations were performed where energy-time and time-bin entangled photon were successfully generated from AlGaAs waveguides~\cite{Sarrafi_2014, Autebert_2016, Chen2018} (Fig.~\ref{fig7}b), albeit with external pumps. 

A very recent demonstration combined an AlGaAs waveguide with a polymer-based integrated photonic circuit containing a long pass filter and a polarizing beam splitter to route the signal and idler photons to two different outputs, which were then connected to optical fibers~\cite{Thiel2024}. One important drawback is the lower non-linear coefficient of AlGaAs compared with other $\chi^{(2)}$ crystals, and thus the brightness of these sources is considerably lower than lithium niobate or KTP for instance. Silicon waveguides have also been used, except in this case, the crystal is a $\chi^{(3)}$ non-linear material, in which the entangled photon pairs are generated from SFWM~\cite{Takesue_2007, Xiong_2015}, which is much less efficient than SPDC. An advantage of working with silicon is due to the integration possibilities with other electronic components, which benefit from the huge electronics industry. Finally, one experiment employed a waveguide in a silicon chip composed of many silicon photonic crystal nanocavities, thus benefitting from a slow-light enhanced SFWM effect, successfully generating time-bin entanglement~\cite{Takesue_2014}.


\subsection*{Micro-ring resonators}


The previous approach began to tackle the integration aspect by making it easier to couple to optical fibers as well as combining multiple non-linear process in the same crystal. Another approach based on integrated photonic circuits takes integration further by opening the possibility to integrate the photon pair source with many other components and even detectors in the same chip~\cite{Wang2020, Elshaari2020, Giordani2023}. 

A direct comparison between the spectral brightness of a straight waveguide and a micro-ring cavity coupled to a waveguide configuration showed an improvement of more than two orders of magnitude, showing the promised potential of micro-ring resonators in integrated photonics for quantum technology applications~\cite{Clemmen2009}. This cavity enhancement is important as it boosts the weak third-order non-linearity $\chi^{(3)}$ SFWM effect. These first experiments focused on the generation of correlated photons improving the brightness and coincident-to-accidental ratio (CAR)~\cite{Azzini2012, Engin2013}. Following these promising results, it was only natural that such integrated sources would be used to successfully generate energy-time~\cite{Grassani_2015, Wakabayashi_2015, Ma_2017} and time-bin~\cite{Reimer_2016, Zhang:18} entanglement. 

As this source design is based on a resonator the photon pairs are emitted in well defined correlated discrete wavelengths covering the emission spectrum dictated by the phase-matching conditions~\cite{Reimer_2016}. With appropriate tuning, this naturally gives rise to the generation of photon pairs distributed across specific wavelength channels within the telecom spectral band in optical fibers, around 1550 nm~\cite{Reimer_2016, Mazeas_2016, Jaramillo-Villegas_2017, Oser_2020, Fan_2023} (Fig.~\ref{fig7}c). With appropriate design of the micro-resonator itself, it is even possible to have a source emitting pairs at both 1310 and 1550 nm spectral bands when dual pumped, allowing more possibilities for integration with telecom optical networks~\cite{Ma_2018}. In this same work, taking advantage of the silicon on insulator platform, a p-i-n monitoring photodiode was included within the ring, to monitor the resonances in real-time. To avoid the issue of having a tunable pump laser, necessary to match the micro-ring resonances, Garrisi \textit{et al.}\ used the ring itself as a filter to an external fiber laser, pumped by a semiconductor optical amplifier~\cite{Garrisi2020}.

Other materials have also been explored to build integrated sources based on micro-ring resonators, such as AlGaAs~\cite{Steiner_2021, Steiner2023} and InP~\cite{Kumar2019}, opening possibilities for integration with light sources such as laser diodes. Other interesting results include the use of an aluminum nitride micro-ring to generate energy-time entangled photons from three input pump photons, effectively being a synthetic $\chi^{(4)}$ process~\cite{Wang2022b}, generation of energy-time and polarization hyper-entanglement on a silicon micro-ring resonator~\cite{Suo2015}, creation of entangled energy-time photons using based on an array of ring resonators on silicon~\cite{Mittal2021} and the first integrated photon pair source based on silicon carbide, where energy-time entanglement demonstration was generated~\cite{Rahmouni2024}. Finally, other integrated designs are also possible, such as for example a silicon microdisk coupled to a waveguide has been employed to produce energy-time entangled photon pairs~\cite{Rogers2016}.


\subsection*{Quantum dots}


Although both types of sources previously described have made great strides towards improving brightness and integration, they still are still bound by a probabilistic generation process, which is present in all earlier sources. That stems from the fact that the probability to generate a photon pair from a pump photon is very low and independent from other pump photons, and as a consequence multi-pair generation is a reality unless low pump powers are employed, which compromises generation rate. A completely different approach for a triggered deterministic source was proposed relying on the radiative decay of a biexciton state in a quantum dot~\cite{Benson2000}. The first demonstrations concentrated on the generation of polarization entanglement, and how to tackle the problem of the ```which-path'' information from the non-degenerate exciton levels for horizontal and vertical polarizations~\cite{Akopian2006, Stevenson2006, Young_2006, Muller2009}. 

\begin{figure*}
 \centering
 \includegraphics[width=1.0\linewidth]{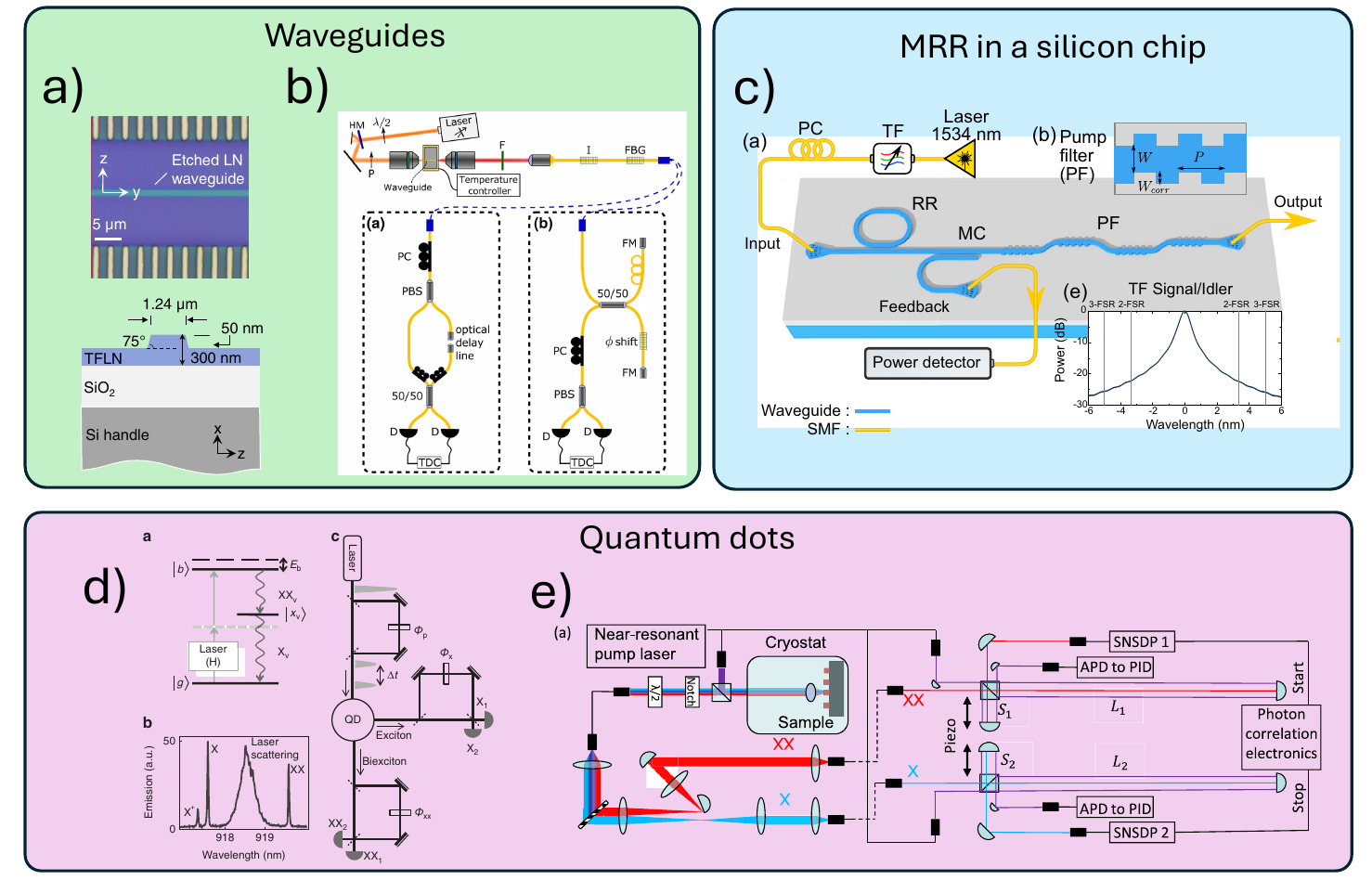}
 \caption{Integrated sources of energy-time and time-bin entanglement. a) Thin-film lithium niobate waveguide source of photon pairs~\cite{Zhao_2020}. Reprinted from \cite{Zhao_2020} Copyright (2020) Authors from a CC BY 4.0 license. b) Energy-time entangled source of photon pairs based on an AlGaAs waveguide~\cite{Autebert_2016}. Reprinted with permission from \cite{Autebert_2016} \copyright (2016) Optical Society of America. c) Micro-ring resonator source of broadband energy-time entangled photons on a Silicon chip~\cite{Oser_2020}. Reprinted from \cite{Oser_2020} Copyright (2020) Authors from a CC BY 4.0 license. d) Generation of time-bin entanglement~\cite{Jayakumar_2014} and e) energy-time entanglement \cite{Hohn2023} from a quantum dot. Reprinted with permission from \cite{Jayakumar_2014} Copyright (2014) Springer Nature and Reprinted from \cite{Hohn2023} Copyright (2023) Authors from a CC BY 4.0 license.}
 \label{fig7}
\end{figure*}

For time-bin generation, the approach has focused on pumping the dot with an early and a late pulse tuned for coherent resonant two-photon excitation~\cite{Jayakumar_2014} (Fig.~\ref{fig7}d). The pulses are optimized to have equal probability of generating the photon pair in either pulse as well as improving state fidelity~\cite{Huber2016}. Here, there is no issue on forcing the exciton levels to be degenerate with the polarization state, since only of one of the two polarization decay paths is used. However, this becomes a probabilistic source, since in order to avoid multi-photon emission, the generation efficiency per pulse must be kept low. 

A proposal to solve this problem was made earlier relying on an additional metastable level, which promotes the quantum dot to the excited biexciton state if absorbed by an appropriate pump pulse~\cite{Simon2005}. The first pulse has a certain probability to promote the system from the metastable level to the biexciton state and then it quickly decays to the ground level emitting a photon pair. The second pulse will not be absorbed as it is out of resonance. If the first pulse fails to induce the transition then the second pulse will, thus creating a time-bin entangled state. This scheme has unfortunately never been performed in practice due to a lack of a suitable metastable level in the quantum dot system employed until today.

A subsequent study created time-bin entangled photons starting with a single polarization entangled pair generated directly from a quantum dot, and converted it to time-bin with an interface comprised of an unbalanced Mach-Zehnder interferometer with polarization beamsplitters~\cite{Versteegh_2015}, thus removing the probabilistic nature of previous efforts. Then a successful demonstration of time-bin and polarization hyperentanglement from a quantum dot was carried out, with the degeneracy of the exciton level to erase the which-path information in the decay paths for polarization entanglement obtained by growth engineering~\cite{Prilmuller_2018}. 

Further work was done in optimizing time-bin photon pair extraction from the quantum dot~\cite{Gines2021} as well as from a quantum dot embedded in a nanowire~\cite{Aumann2022}. In~\cite{Sun2017b}, Franson interferometry was used to characterize the inhomogeneous broadening of the bi-exciton state in a quantum dot, although the visibility was far too low to violate a Bell inequality. Later, a successful energy-time Bell violation was shown directly from a quantum dot emission~\cite{Hohn2023} (Fig.~\ref{fig7}e), the only result of this type so far from this type of source. Finally, energy-time entanglement was also generated from inelastic scattering off a quantum dot coupled to a waveguide \cite{Liu2024} and from a Rubidium atom coupled to a fiber Fabry-P\'erot cavity \cite{Wang2025}.

\section{Applications}


Over the years many applications that benefitted directly from energy-time or time-bin entanglement have been proposed and demonstrated. Here we summarize key results in diverse areas of quantum information processing. 


\subsection*{Quantum key distribution}


The most well-known application of long-distance entanglement distribution is arguably the generation of cryptographic keys between remote parties~\cite{Ekert:1991PRL, Ekert1992, BBM92}. The first demonstration employed energy-time entanglement to distribute keys over optical fibers with the two photons generated at the 1300 nm window~\cite{Tittel2000}. Furthermore, this experiment employed the so-called time and energy mutually unbiased bases~\cite{Bechmann-Pasquinucchi2000PRA}, where Alice and Bob on the first case check whether their corresponding photon arrived on a side peak, or in the second basis, measure which interferometer output was generated in the central time peak. This demonstration was later improved to non-degenerate photons at 810 and 1550 nm, with the later one propagating over an 8.5 km fiber channel to Bob~\cite{Ribordy2000}, and then later to 31 km~\cite{Fasel2004}. 

Then, with the use of superconducting detectors, plane light wave circuits implementing the unbalanced interferometers, as well as a high 1 GHz repetition rate for the pump pulse, a successful demonstration of QKD over a total distance of 100 km over dispersion-shifted optical fibers was performed~\cite{Honjo2008} (Fig.~\ref{fig8}a). An interesting proposal that does not rely on entanglement was also made, which actually employs Franson interferometry and a central single-photon source emitting two consecutive photons to carry out quantum key distribution~\cite{Inoue2003}. 

More recently the trend has shifted to multi-user QKD distribution. This is based on a broadband source of entangled photons, which are then distributed to multiple users, each one allocated a specific spectral channel, usually matched to the dense wavelength division multiplexing (DWDM) standard channels employed in fiber-optic communications~\cite{Wengerowsky2018}. This scheme naturally matches a star network configuration, with the source deployed as the central node. This first experiment employed polarization entanglement, distributing keys across multiple users in a DWDM network. 

A first demonstration in this direction with energy-time entanglement had already taken place in 2016, albeit in a proof-of-principle experiment using tunable filters~\cite{Kaiser_2016}. Then, by taking advantage of an integrated micro-ring resonator source built on silicon nitride, energy-time entangled photon pairs were generated across 6 wavelength pairs, and routed using DWDM filters to four different users~\cite{Wen2022}, such that keys can be automatically generated between any user-pair combination. 

The resonator configuration helps to minimize filtering losses as the ring resonances can be designed to match the DWDM channels. Then, a broadband time-bin source based on SPDC combined with SHG was built and photon pairs distributed to four users also matched to DWDM channels, and this time with one of the users connected over a deployed fiber loop of 26.8 kms, and the other users separated to the source with fiber spools up to 81.2 km~\cite{Fitzke2022}. 

A similar experiment within a star network was performed where time-bin entangled photon pairs were distributed among 3 users in order to generate secret keys among them in any pairwise combination, where each user pair was separated by 60 km of optical fibers~\cite{Kim2022}. Finally in~\cite{Mueller2024} (Fig.~\ref{fig8}b) a high-repetition rate time-bin source (pump clocked at 4.09 GHz), high visibility (higher than 99\%) over multiple DWDM channel pairs was demonstrated with the photon pairs generated using SHG and then SPDC processes.


\subsection*{Entanglement-assisted communication}


Other communication tasks providing quantum advantages have been demonstrated using energy-time or time-bin entanglement. A very early proposal is non-local dispersion compensation, where the dispersion in time suffered on correlated photons as they propagate over dispersive media, such as optical fibers, can be compensated if negative dispersion is applied to only one of the photons from the pair and they are both measured in coincidence non-locally~\cite{Franson1992}. 

A practical metrological application was demonstrated in 1998, where the chromatic dispersion of optical fibers was measured around 1300 nm, although both photons traveled through the same optical fiber~\cite{BRENDEL1998}. Another demonstration following the original proposal was performed in~\cite{Baek2009} with non-degenerate down-converted photon pairs (750 and 896 nm). An inequality was derived to verify entanglement through non-local cancellation of dispersion~\cite{Wasak2010}, and then later experimentally tested~\cite{Li2019}. 

Other studies also concentrated on dispersion cancellation on Hong-Ou-Mandel interference, although the photon pairs are measured locally in this case~\cite{Steinberg1992, Resch2007, Ryu2017}. Another interesting aspect is that non-local dispersion (NLD) cancellation has been used as a tool to compensate dispersion over optical fiber channels in order to improve the QKD rates of a polarization entanglement-based setup~\cite{Neumann2021}. Finally, a proposal was recently made extending NLD cancellation to three or more photons~\cite{Nodurft2020}, albeit interestingly it is not possible to fully cancel it, as in the two-photon case. 

Other well-known examples of quantum communication protocols that require entanglement are quantum teleportation~\cite{Bennett_teleportation} and entanglement swapping~\cite{Zukowski_1993}. In the first, a joint projection between an unknown quantum state $|\Psi_{in}\rangle$ together with a photon from an entangled pair is made onto a Bell state. This operation is called a Bell state measurement (BSM), and in the qubit case, the projection is done into one of the maximally entangled Bell states $|\phi^{\pm}\rangle = 1/\sqrt{2}(|00\rangle \pm |11\rangle)$ and $|\psi^{\pm}\rangle = 1/\sqrt{2}(|01\rangle \pm |10\rangle)$. The output of this projection produces two bits of classical information, which define which out of four possible rotation operations to apply to the other photon of the entangled pair in order to recover the unknown state $|\Psi_{in}\rangle$. If the whole operation is successful we have $|\Psi_{out}\rangle = U|\Psi_{in}\rangle$, where $U = \{I, X, Y, Z\}$, corresponding to the identity and one of the three Pauli matrices, respectively~\cite{Bennett_teleportation}. 

Entanglement swapping is similar in concept to teleportation, but with the main difference that an entangled state is itself teleported instead of a single quantum state. In this case, a BSM is applied to two photons each belonging to a different photon pair, and then entanglement is created between the other two photons. These protocols are key elements needed in the use of quantum communications in an interconnected ``Quantum Internet''~\cite{Wehner2018}. 

The first experimental demonstrations of quantum teleportation were done over short distances on an optical table based on polarization~\cite{Bouwmeester1997}, and polarization-path entanglement~\cite{Boschi1998}. Due to the impossibility of a BSM based on linear optics distinguishing between all four Bell states, teleportation demonstrations typically rely on a partial BSM, where only the states $|\Psi^{\pm}\rangle$ can be distinguished. 

The first experimental demonstration of quantum teleportation of a time-bin qubit was also the first one to demonstrate real physical separation between Alice, who prepares the state $|\Psi_{in}\rangle$, and Bob who receive it~\cite{Marcikic2003}. In this experiment Charlie, who performs the BSM between Alice's state and one of the photons from the entangled pair, is also located in Alice's lab as well as the source of photon pairs. Alice and Bob are spatially separated by 55 m and connected with 2 km long optical fiber spools. Time-bin entangled photons are generated using SPDC with the pump pulses prepared with an interferometer. 

A critical element of teleportation is the interference of independent and identical photons~\cite{Riedmatten2003} from different sources within Charlie who performs the BSM. In the experiment described here~\cite{Marcikic2003} this was done by splitting the consecutive pump pulses into two paths, and feeding these each through an independent non-linear crystal, both designed to produce time-bin entangled photon pairs at the two telecom windows of 1310 and 1550 nm. Alice prepares $|\Psi_{in}\rangle$ on the 1310 nm photon from one crystal using an unbalanced time-bin interferometer. This photon interferes on a beamsplitter with the other 1310 nm photon from the other crystal performing the BSM operation. The other photon at 1550 nm propagates through the 2 km optical fiber channel (55 m of spatial separation) to Bob in a different lab, who has an identical time-bin interferometer as Alice, to perform the qubit analysis conditioned on a double detection event in the BSM happening on the two opposing detectors with a time difference equal to the unbalance in Alice and Bob's time-bin analyzers, corresponding to a successful projection onto the $|\Psi^-\rangle$ state. These results showed the first long-distance quantum teleportation experiments.

Teleportation is also a key element of quantum repeaters, which aim to increase the transmission distance of quantum communication beyond the limit dictated by the noise level of the single-photon detectors~\cite{Briegel1998}. A simpler version, which does not require quantum memories and entanglement purification, is referred to as a quantum relay~\cite{Jacobs2002}. The previous time-bin teleportation demonstration~\cite{Marcikic2003} was shortly afterwards expanded into a relay configuration, by adding a 2 km fiber spool between Alice and the BSM operation at Charlie~\cite{Riedmatten2004}. 

Then the BSM operation was improved to probabilistic distinguishability of 3 out of the 4 Bell states with a Bell-state analyzer that comprised of an unbalanced Mach-Zehnder interferometer, having the same delay between two consecutive time-bins and one photon entering from each input port in the first beamsplitter~\cite{Houwelingen2005, Houwelingen2006}. The first time-bin teleportation demonstration in a deployed configuration took place in 2007 with the laboratory containing Alice and Charlie located 550 m away from Bob, connected through 800 m of installed optical fiber~\cite{Landry2007}. 

A major challenge of these experiments without a quantum memory, is that the arrival time of the photons at the BSM needs to be precisely controlled over long distances. An experiment relaxed this constraint by resorting to energy-time entanglement combined with ultra-narrow spectral filtering~\cite{Halder2007}, successfully demonstrating energy-time entanglement swapping, also discussed as a proposal in~\cite{Messina2007}. 

\begin{figure*}
 \centering
 \includegraphics[width=1.0\linewidth]{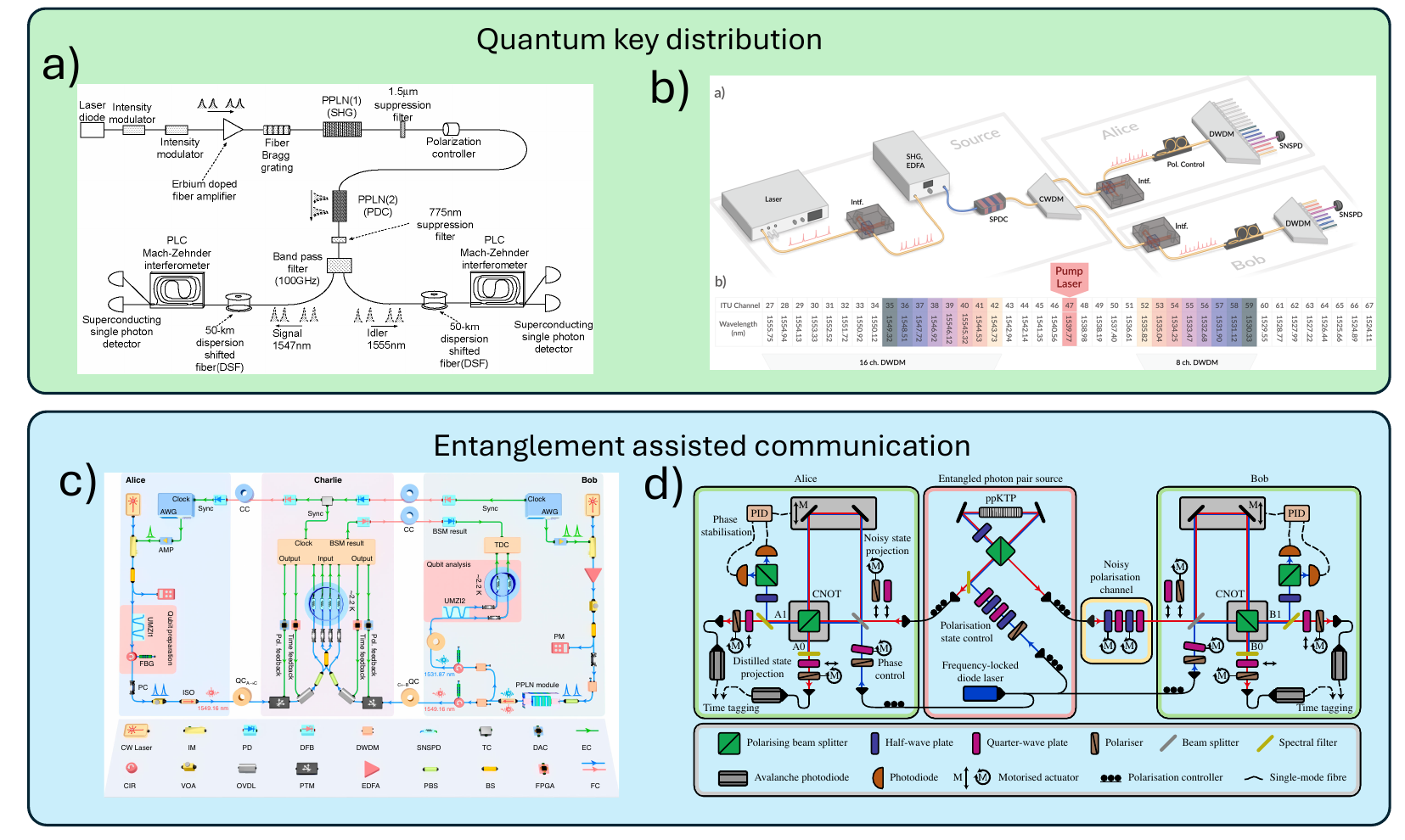}
 \caption{Entanglement-assisted quantum communication applications. a) Quantum key distribution with time-bin entangled photons over 100 km distance through spooled optical fibers~\cite{Honjo2008}. Reprinted with permission from \cite{Honjo2008} \copyright (2008) Optical Society of America. b) High-rate multi-user quantum key distribution based on time-bin entanglement~\cite{Mueller2024}. Reprinted with permission from \cite{Mueller2024} \copyright (2024) Optica Publishing Group. c) High-rate quantum teleportation over metropolitan optical fiber links~\cite{Shen2023}. Reprinted from \cite{Shen2023} Copyright (2023) Authors from a CC BY 4.0 license. d) Entanglement purification using polarization/energy-time hyper-entanglement~\cite{Ecker2021}. Reprinted with permission from \cite{Ecker2021} Copyright (2021) American Physical Society.}
 \label{fig8}
\end{figure*}

As with other quantum communication protocols, advances in technology greatly improved the performance of later experiments. In 2015, time-bin entanglement teleportation over 100 km of optical fibers was carried out with the advent of superconducting single-photon detectors~\cite{Takesue2015}. Then two parallel experiments demonstrated teleportation over a metropolitan optical network~\cite{Sun2016, Valivarthi2016} and another one entanglement swapping~\cite{Sun2017}. Finally, more recent results managed to obtain teleported state fidelities of $> 90\%$ over 22 km of optical fibers~\cite{Valivarthi2020}, a teleportation rate of 7 Hz over 64 km of deployed fiber~\cite{Shen2023} (Fig.~\ref{fig8}c) and multi-user entanglement swapping~\cite{Li2023}. 

Other communication tasks based on energy-time and time-bin entanglement have also been carried out. Very early on, quantum secret sharing was performed with time-bin entanglement~\cite{Tittel2001}. The authors did not employ true three-partite Greenberger-Horne-Zeilinger states, but instead noted that employing the pump photon as the third party they could mimic the same correlations. A much more recent experiment demonstrated a network of interconnected 15 users, performing quantum secure direct communication between them, by employing sum frequency generation to deterministically distinguish between all four Bell states~\cite{Qi2021}. 

Another example is superdense coding that was performed over short optical fiber links, together with complete Bell state measurements relying on hyperentanglement between polarization and energy-time~\cite{Williams2017}. Finally, entanglement purification \cite{Sheng2014, Yan2021, Yan2023} has been successfully performed, once again relying on polarization/energy-time hyperentanglement, but in this case to realize the required CNOT operation as well as lowering the requirement to only one copy, thus requiring only one photon pair~\cite{Ecker2021} (Fig.~\ref{fig8}d).


\subsection*{Higher dimensions}


Expanding the Hilbert space of a quantum system 
beyond the qubit case brings in new possibilities in quantum information processing~\cite{Erhard2020, Cozzolino2019}. A straightforward advantage is the encoding of more information per photon as the dimensionality increases, which also improves the noise tolerance in QKD~\cite{Cerf2002}. Higher dimensions also allows some tasks in quantum information that are not possible when one is limited to qubits~\cite{Budroni:2022RMP}, as well as reducing the required detection efficiency for loophole-free Bell tests~\cite{Miklin2022, Xu2023}. Mathematically, a general $d$-dimensional qudit is expressed simply as $|\Psi\rangle = 1/\sqrt{d}\sum_d \alpha_d e^{i\phi_d}|d\rangle$, where $\alpha_d$ are real coefficients, such that $\sum\alpha_d^2 = 1$, $\phi_d$ are relative phases, and $|d\rangle$ are orthogonal states. Polarization is limited to a bi-dimensional space, but the spatial~\cite{Canas2017}, frequency~\cite{Imany2018} and time-bin/energy-time~\cite{Islam2017} degrees of freedom are all suitable for high-dimensionality. 

First attempts focused on the generation of high-dimensional time-bin entanglement when a non-linear SPDC crystal is pumped by a train of coherent pulses~\cite{Riedmatten2002, Riedmatten2004b}. Then, a Bell test, using the CGLMP-Bell inequality~\cite{Collins2002}, was performed on energy-time entangled qutrits, generated from continuously pumping an SPDC non-linear crystal, and employing three-arm unbalanced MachZenhder Franson-like interferometers as the analyzers~\cite{Thew2004} (Fig.~\ref{fig9}a). 

A major difficulty expanding to higher-dimensional time-bin and energy-time entanglement is the difficulty in building stable multi-arm interferometers. One solution is to employ cascaded two-arm interferometers with varying delays~\cite{Islam2017, Richar2012}. One study showed that such cascaded interferometers are important for the security of high-dimensional QKD protocols based on energy-time entanglement~\cite{Brougham2013}. Another study showed that high-dimensional entanglement can be characterized using only two output interferometers, applying the technique for time-bin entanglement~\cite{Martin2017}. Recently another approach is based on the use of multi-mode fibres to carry out generalised projective measurements for high-dimensional time-bin entanglement \cite{Srivastav2024}.

Another approach takes advantage of the fact that measurement in the time-bin or energy-time computational basis (time basis) is straightforward, as it is simply the time of arrival compared to a common reference between Alice and Bob. In the other basis of typical interest (the energy basis), composed of a linear superposition of the computational basis states, a compromise has been the use of two-arm unbalanced interferometers with variable length for the long arm. In this way, all double-state superposition combinations of the energy basis can be tested. 

This concept applied to energy-time entanglement was used to demonstrate one of the first high-dimensional QKD experiments~\cite{Ali-Khan2007}, with a more recent experiment also demonstrating quantum steering \cite{Chang2024}. A security proof was then derived employing an analogy to continuous-variable (CV) entanglement, when dual-basis interferometry is used, alternating between Franson's interferometers and its frequency conjugate~\cite{Zhang2014}. A more practical proof considering decoy states to tackle multi-pair emission is presented in~\cite{Bunandar2015}, where a variation of a setup generating energy-time entanglement is employed where the measurements are performed in the conjugate time and frequency bases. 

An experimental implementation for high-dimensional energy-time entanglement QKD is then carried out where the time basis is used to generate the correlated key bits with a dimensionality of 1000, and the energy basis with Franson interferometers is used to check security~\cite{Zhong2015}. A very recent demonstration combined an SFWM source in integrated photonics, with the analyzers also implemented in photonic chips with unbalanced cascaded Mach-Zehnder interferometers generating up to 8-dimensional time-bin entanglement, and entanglement-based QKD over up to 60 km propagation over optical fibers~\cite{Yu2025}. Finally, recent efforts have also focused on the characterization of spectral Franson interference for high-dimensional applications \cite{Jin2024b}.

An alternate technique to increase the dimensionality, which does not have the issue of scaling up or cascading the measurement interferometers, is to combine energy-time or time-bin qubit entanglement with other degrees of freedom, the so-called hyper-entanglement~\cite{CinelliPRL2005, Barreiro2005}. This is a popular alternative to increase the dimensionality as opposed to multi-partite entanglement, where more qubits of the same type are employed in the same system~\cite{Horodecki_entanglement}. A hyperentangled state is one that can be written in the following form: $|\Psi_{\textrm{HE}}\rangle = |\Phi_{a}\rangle \otimes |\Phi_{b}\rangle \otimes |\Phi_{c}\rangle \otimes \ldots $, where $|\Phi_{i}\rangle$ corresponds to a two-qubit pair encoded in degree of freedom $i$. An advantage of this approach is that one avoids the scaling of multi-photon coincident detection when bound by losses and limited detection efficiency. Hyperentanglement sources based on energy-time entanglement as one of the degrees of freedom 
are quite straightforward as 
this form of entanglement stems directly from any source with continuous pumping. 

First implementations of hyper entangled states using energy-time entanglement as one of the degrees of freedom were carried out in~\cite{Barreiro2005} and then in~\cite{Vallone2009}, by combining polarization and spatial modes together with energy-time. Through the use of an external cavity, hyperentanglement between energy-time, frequency-bin and polarization was also achieved~\cite{Xie2015,Chang2021}. Based on broad-band entanglement emission, a source based on polarization energy-time hyperentanglement was demonstrated with the emission bandwidth compatible with DWDM telecom channels~\cite{Vergyris2019}. With a focus on high-dimensional communication, hyperentangled states between polarization and energy-time were distributed over a 1.2 km free-space link~\cite{Steinlechner2017}. Moving away from communication applications, the creation of cluster states for one-way quantum computing was realized employing hyperentangled states between time and frequency-bin~\cite{Reimer2019}. 

Other experiments have focused on the advantages given by high-dimensional entanglement. In~\cite{Ecker2019}, it was explicitly demonstrated that both energy-time and orbital angular momentum (OAM) higher-dimensional entanglement is more robust to noise. Specifically, as a function of the noise intensity, an entanglement witness could only be violated for high-dimensional entanglement. Another experiment used hyper-entanglement between polarization and frequency-time generated from an SPDC source to filter out uncorrelated random added noise using the frequency (essentially energy) correlations, while measuring the information content on the polarization information~\cite{Kim2021}. Robustness to noise due to high-dimensional energy-time entanglement was then explored in a 10.2 km free-space link in Vienna, where it was explicitly shown that communication under rainy conditions and under the presence of sunlight was only possible when employing high-dimensionality~\cite{Bulla2023}. Finally, in a very recent study, the authors demonstrated improvements in secret key rate when employing high-dimensional hybrid polarization time-bin/polarization entanglement compared to the standard bi-dimensional case, up to a 50 km propagation distance over spooled optical fibers~\cite{Zhong2024} (Fig.~\ref{fig9}b).

In spite of its difficulties, tri-partite energy-time entanglement was actually generated by taking advantage of two SPDC cascaded processes, where one of the generated photons in the first process acts as a pump in the second \cite{Shalm2012}. This technique was then successfully used to demonstrate three-photon non-local interference with the individual propagation paths all placed in the same unbalanced Michelson interferometer for extra stability \cite{Agne2017}. In spite of the low creation of three-photon events, this technique has the advantage of not needing to post-select a three-photon state out of a four photon state produced by coherent SPDC processes.


\subsection*{Entanglement storage}


Key elements for future quantum networks are quantum repeaters, which perform multiple in-line entanglement swapping operations in order to extend the transmission distance~\cite{Briegel1998, Duan2001}. Quantum repeaters are dependent on memories to be able to synchronize the joint measurement operations performed at each hop, toward the construction of a fully functioning Quantum Internet~\cite{Kimble2008, Wehner2018}. 

An early demonstration of entanglement storage showed a very small delay applied to one of the photons of an energy-time entangled pair, when passing through a Rubidium vapour cell through the slow light effect~\cite{Broadbent2008}, showing that in spite of some degradation, the two-photon fringe visibility was still sufficient to violate a Bell inequality. However, for a practical applications more flexibility would be needed in terms of storage times and on-demand retrieval. 

One promising technique that was proposed is the atomic frequency comb (AFC)~\cite{Afzelius2009}, where the absorption shape of an atomic ensemble is tuned into a comb using optical pumping, allowing the quantum state to be stored. It is then retrieved when the atoms in this comb structure rephase, which depends on the comb period. If on-demand retrieval is desired the state can be stored in an appropriate ground state spin level with the help of an additional control laser, and later retrieved back at will with a second counter-propagating control laser. Appropriate atomic systems are rare-earth ions as they allow the formation of high-resolution combs with coherence times of $\mu$s to ms~\cite{Afzelius2009}. 

\begin{figure*}
 \centering
 \includegraphics[width=1.0\linewidth]{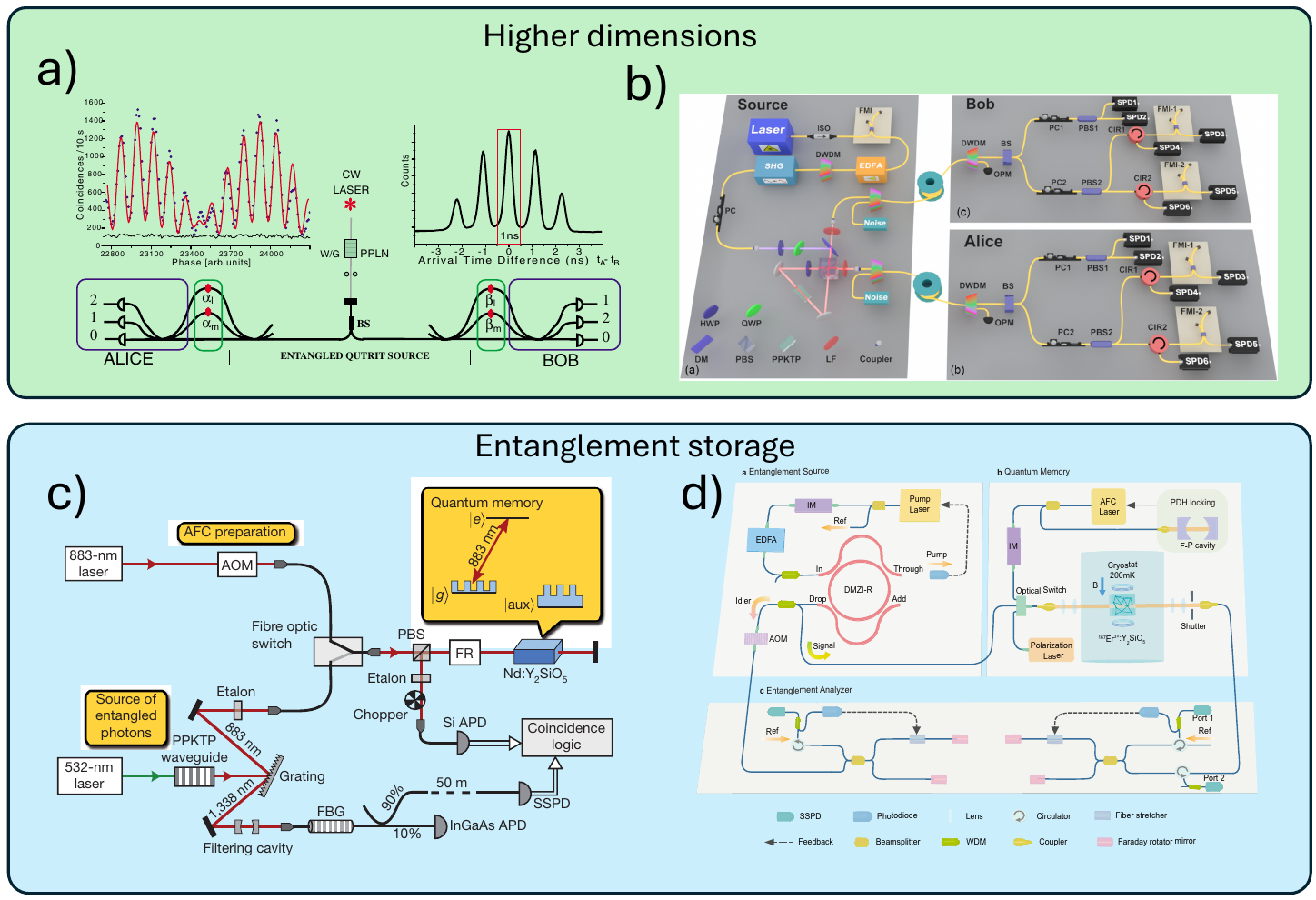}
 \caption{High-dimensional entanglement and energy-time/time-bin entanglement storage. a) High-dimensional time-bin Bell tests with qutrits~\cite{Thew2004}. Reprinted with permission from \cite{Thew2004} Copyright (2004) American Physical Society. b) Distribution of hyper-entanglement time-bin/polarization entanglement over 50 kms of optical fibers~\cite{Zhong2024}. Reprinted with permission from \cite{Zhong2024} \copyright (2024) Optica Publishing Group. c) Storage of time-bin entanglement on a neodymium-doped crystal~\cite{Clausen2008} and d) an Erbium-doped crystal~\cite{Jiang2023}, both using the AFC protocol. Reprinted with permission from \cite{Clausen2008} Copyright (2008) Springer Nature and Reprinted from \cite{Jiang2023} Copyright (2023) Authors from a CC BY 4.0 license.} 
 \label{fig9}
\end{figure*}

Storage of one of the photons of an energy-time entangled pair was successfully demonstrated using the AFC protocol where the memory consisted of a crystal doped with neodymium ions supporting the storage of 883 nm photons~\cite{Clausen2008} (Fig.~\ref{fig9}c). The other photon from the pair (1338 nm) was propagated over 50 m, and then joint measured with the 883 nm photon released from the memory. A successful CHSH Bell inequality violation was shown following storage. 

A follow-up experiment based on the same ion showed successful storage of a photon from a polarization/energy-time hyperentangled photon pair~\cite{Tiranov2015}. The memory crystal consisted of two separate identical crystals separated by a thin half-wave plate. The same group also showed the quantification of multidimensional energy-time entanglement with one of the photons stored in the same type of memory when only having access to partial data~\cite{Tiranov2017}. 

Another experiment targeted erbium ions as the atomic system to implement storage through the AFC scheme~\cite{Saglamyurek2015}. The main advantage of erbium is the available resonance at 1532 nm, which is directly compatible with the lowest attenuation spectral region in telecommunication optical fibers, making such a memory compatible with long-distance networks. Another advantage was the use of available commercial Erbium doped fibers, thus facilitating integration into optical networks. Unfortunately, the maximum storage time was very short (35 ns), owning to the low coherence time of the Erbium ions due to their placement in the amorphous structure of the optical fiber. 

Another employed rare-earth ion is praseodymium, which was first used in a scheme to generate entanglement in time between a photon and a spin collective state in a crystal doped with the ion~\cite{Kutluer2019}. Then successful storage of a photon from an energy-time entangled pair was performed in the same memory type, showing both AFC and spin-wave storage (on-demand)~\cite{Rakonjac2021}. Then an improvement was done where a waveguide was laser-written on the crystal, and then integrated directly to an input and an output optical fiber, improving the overall memory efficiency~\cite{Rakonjac2022}. Finally, a recent paper showed the storage of a photon from an energy-time entangled pair using the AFC protocol on erbium ions doped in a crystal, showing considerable improvements in storage time compared to what was obtained before with the same ion~\cite{Jiang2023} (Fig.~\ref{fig9}d). The authors also employed a micro-resonator in a silicon-nitride integrated photonics chip as the source of photon pairs, which were both naturally generated within the telecom spectral band owning to the spontaneous four-wave mixing generation process. 

\section{Outlook}

Energy-time and time-bin entanglement have been instrumental in advancing quantum information science as well as our understanding of quantum foundations. Although energy-time entanglement stems directly from energy-conservation in all processes of photon-pair generation, and thus 
relatively easy to obtain, the measurement process is more complicated due to the interferometric requirements. Therefore, compared to polarization entanglement, energy-time entanglement is easier to produce but more complicated to characterize. 

Its success came from its increased robustness when propagating over optical fiber channels, which showed for the first time that entanglement could be maintained over long distances. The creation of time-bin sources, where the pump is modified to be a coherent superposition of pulses, extended the transmission distance even further, due to the possibility of a distributed synchronisation signal. 

Early experiments focusing on the generation of photon pairs using bulk spontaneous parametric down-conversion, which although highly successful, have been overshadowed nowadays by more efficient or integrated processes, such as periodically poled waveguided crystals, quantum dots and integrated resonators. 
Apart from pure point-to-point communication tasks, such as QKD and entanglement distribution, many other important schemes were demonstrated such as multi-user communication, storage of quantum information into memories and high-dimensional quantum information processing. 

Nevertheless, in spite of many advances, there are many challenges to pursue. One clear gap to cover is improving and expanding the demonstrations removing the post-selection loophole, opening the path for stronger fundamental tests as well as towards device-independent quantum communication. Advances in integrated photonics~\cite{Wang2020} should make it easier to remove the loophole, due to the possibility of implementing complex and stable interferometers and switches on chip. 

One important start in this direction was the recent implementation of the hug interferometer on a silicon nitride chip~\cite{Santagiustina2024}. Another relatively unexplored area is higher dimensions, which would open up the path for more 
ambitious protocols, as well as direct advantages such as more information encoded per particle. In this respect, the recent advances on integrated photonics and sources make up a robust framework for expanding this major field in the future, as was very recently seen in~\cite{Yu2025}. It is interesting to note that energy-time and time-bin entanglement are the current forms of entanglement most suited to support both high-dimensionality and long-distance propagation simultaneously, key traits for future quantum communication schemes. Combined with the development of multi-dimensional quantum memories, the future for energy-time and time-bin quantum communications is certainly bright. 

\textit{Note:} During the preparation of this manuscript we became aware of a separate review on the topic of time-bin quantum states with a different focus~\cite{Yu2024review}.


\section{Acknowledgements}
The authors would like to thank Yu-Bo Sheng, Gui-Lu Long, Jian Wang, Michael Lubasch, Kai-Chi Chang, Mehul Malik and Zi-Qi Zeng for helpful discussions. This work is part of the project SECuRe quantum communication based on Energy-Time entanglement (SECRET), QuantERA Call 2019. We wish to thank Zenith Link\"{o}ping University, the QuantERA grant SECRET (VR 2019-00392), VR regular grant (VR 2023-05031), the EU-funded project FoQaCiA, the MCINN/AEI (Project No.\ PID2020-113738GB-I00) for financial support. 



\bibliography{common}


\end{document}